\documentclass[useAMS,usenatbib]{mn2e}
\usepackage{graphicx}
\usepackage{epsfig}
\usepackage{rotating}
\usepackage{amssymb}
\usepackage{color}

\usepackage{fancyhdr}
\usepackage{textcomp}
\usepackage{longtable}
\usepackage{tabularx}
\usepackage[bookmarks]{hyperref}

\def\um{$\textrm{$\mu$m}\ $}                             % microns \um
\def  \cmsq     {\ifmmode {\rm cm}^{-2} \else cm$^{-2}$\fi}
\def  \ergs     {\ifmmode {\rm erg\,s}^{-1} \else erg s$^{-1}$\fi}
\def  \ergcms   {\ifmmode {\rm erg\,cm}^{-2}\,{\rm s}^{-1}
                        \else erg\,cm$^{-2}$\,s$^{-1}$\fi}
\def \lhard  {\ifmmode {\rm L_{2-10keV}} \else ${\rm L_{2-10keV}}$\fi}
\def \lir  {\ifmmode {\rm L_{IR}} \else ${\rm L_{IR}}$\fi}
\def \nh  {\ifmmode {\rm N_{H}} \else ${\rm N_{H}}$\fi}
\def \Msun {\ifmmode M_{\odot} \else $M_{\odot}$\fi}
\def \Lsun {\ifmmode L_{\odot} \else $L_{\odot}$\fi}

\def   \sun	{\odot}	
\def \rvir	 {\ifmmode R_{200} \else $\rm R_{200}$\fi}

\title[SFR--density relation up to $z\sim1.6$]{Reversal or no reversal: the evolution of the star formation rate--density relation up to $z\sim1.6$}
\author[F. Ziparo et al.]{F. Ziparo$^{1,2}$\thanks{E-mail:
fziparo@star.sr.bham.ac.uk}, P. Popesso$^{1}$, A. Finoguenov$^{1, 3, 4}$, A. Biviano$^{5}$, S. Wuyts$^{1}$, D. Wilman$^{1}$, 
\newauthor
M.~Salvato${^1}$, M.~Tanaka${^6}$, K.~Nandra$^{^1}$, D.~Lutz$^{1}$, D.~Elbaz $^{7}$, M.~Dickinson$^{8}$,  
 \newauthor   B.~Altieri$^{9}$, H. Aussel$^{7}$, S. Berta$^{1}$, A. Cimatti$^{10}$, D. Fadda$^{11}$, R. Genzel${^1}$, E. Le Floc'h$^{7}$,  
 \newauthor  B. Magnelli$^{12}$, R. Nordon$^{13}$, A. Poglitsch$^{1}$, F. Pozzi$^{10}$ M. Sanchez Portal$^{9}$, L. Tacconi$^{1}$ 
 \newauthor F. E. Bauer$^{14,15}$, W. N. Brandt$^{16}$, N. Cappelluti$^{17, 4}$, M. C. Cooper$^{18}$, J. S. Mulchaey$^{19}$\\
$^{1}$Max-Planck-Institut f\"{u}r extraterrestrische Physik, Giessenbachstra\ss e 1, 85748 Garching bei M\"{u}nchen, Germany\\
$^{2}$School of Physics and Astronomy, University of Birmingham, Edgbaston, Birmingham B15 2TT, UK\\
$^{3}$Department of Physics, University of Helsinki, Gustaf H\"allstr\"omin katu 2a, 00014 Helsinki, Finland \\
$^{4}$University of Maryland Baltimore County, 1000 Hilltop circle, Baltimore, MD 21250, USA \\
$^{5}$INAF/Osservatorio Astronomico di Trieste, via G. B. Tiepolo 11, 34143 Trieste, Italy\\
$^{6}$ National Astronomical Observatory of Japan, 2-21-1 Osawa, Mitaka, Tokyo 181-8588, JAPAN \\
$^{7}$Laboratoire AIM, CEA/DSM-CNRS-Universit{\'e} Paris Diderot, IRFU/Service d'Astrophysique,  B\^at.709, CEA-Saclay, 91191 Gif-sur-Yvette \\
Cedex, France.\\
$^{8}$National Optical Astronomy Observatory, 950 North Cherry Avenue, Tucson, AZ 85719, USA\\
$^{9}$Herschel Science Centre, European Space Astronomy Centre, ESA, Villanueva de la Ca\~nada, 28691 Madrid, Spain\\
$^{10}$Dipartimento di Astronomia, Universit{\`a} di Bologna, Via Ranzani 1, 40127 Bologna, Italy.\\
$^{11}$NASA Herschel Science Center, Caltech 100-22, Pasadena, CA 91125, USA\\
$^{12}$Argelander-Institut f\"ur Astronomie, Universit\"at Bonn, Auf dem H\"ugel 71, 53121 Bonn, Germany \\
$^{13}$School of Physics and Astronomy, The Raymond and Beverly Sackler Faculty of Exact Sciences, Tel Aviv University, Tel Aviv 69978, Israel \\
$^{14}$Instituto de Astrof\'{\i}sica, Facultad de F\'{i}sica, Pontificia Universidad Cat\'{o}lica de Chile, 306, Santiago 22, Chile \\
$^{15}$Space Science Institute, 4750 Walnut Street, Suite 205, Boulder, Colorado 80301 \\
$^{16}$Department of Astronomy and Astrophysics, 525 Davey Laboratory, The Pennsylvania State University, University Park, PA 16802 \\
$^{17}$INAF-Osservatorio Astronomico di Bologna, Via Ranzani 1 40127 Bologna, Italy \\
$^{18}$Center for Galaxy Evolution, Department of Physics and Astronomy, University of California, Irvine, 4129 Frederick Reines Hall Irvine,\\
CA 92697 USA \\
$^{19}$The Observatoires of the Carnegie Institution of Science, 813 Santa Barbara Street, Pasadena, CA 91101, USA
}

\begin{document}

\date{Accepted 2013 October 3.  Received 2013 September 22; in original form 2013 April 29}

\pagerange{\pageref{firstpage}--\pageref{lastpage}} \pubyear{2002}

\maketitle

%\clearpage

\label{firstpage}

\begin{abstract}
We investigate the evolution of the star formation rate (SFR)--density relation in the Extended Chandra Deep Field South (ECDFS) and the Great Observatories Origin Deep Survey (GOODS) fields up to $z\sim 1.6$.  
In addition to the ``traditional method'', in which the environment is defined according to a statistical measurement of the local galaxy density, we use a "dynamical" approach, where galaxies are classified according to three different environment regimes: group, "filament-like", and field.
Both methods show no evidence of a SFR--density reversal.
Moreover, group galaxies show a mean SFR lower than other environments up to $z\sim 1$, while at earlier epochs group and field galaxies exhibit consistent levels of star formation (SF) activity. 
We find that processes related to a massive dark matter halo must be dominant in the suppression of the SF below $z\sim 1$, with respect to purely density-related processes.
We confirm this finding by studying the distribution of galaxies in different environments with respect to the so-called Main Sequence (MS) of star-forming galaxies. 
Galaxies in both group and ``filament-like'' environments preferentially lie below the MS up to $z\sim 1$, with group galaxies exhibiting lower levels of star-forming activity at a given mass. At $z>1$, the star-forming galaxies in groups reside on the MS. 
Groups exhibit the highest fraction of quiescent galaxies up to $z\sim 1$, after which group, ``filament-like'', and field environments have a similar mix of galaxy types. We conclude that groups are the most efficient locus for star-formation quenching. Thus, a fundamental difference exists between bound and unbound objects, or between dark matter haloes of different masses.

\end{abstract}


%----------------------------------------------------------------------------------------
\section{Introduction}
%----------------------------------------------------------------------------------------

The properties of galaxies in the local Universe appear to depend strongly on their environment. This issue was highlighted by \cite{Dressler_1980} with the so-called morphology--density relation. Namely, massive ellipticals and S0 galaxies are preferentially found in crowded regions, such as cluster cores, while spiral and disk galaxies prefer less dense environments. 

It is also well established that a rather tight correlation exists between morphological type and level of star formation (SF) activity. In general, disk galaxies tend to have a higher SF rate (SFR) than spheroidal systems. Recently, the nature of this relation has been carefully studied up to  $z\sim 2.5$ by \cite{Wuyts2011}, through the use of the deep {\it{Herschel\footnote{{\it Herschel}  is an ESA space observatory with science instruments provided by European-led Principal Investigator consortia and with important participation from NASA.}}} surveys  and well-calibrated complementary SFR indicators on the major blank fields, such as the GOODS (Great Observatories Origin Deep Survey; \citealt{Giavalisco2004}) and COSMOS (Cosmological Evolution Survey; \citealt{Scoville2007}) fields. This work highlights that the so-called Main Sequence (MS) of star-forming systems, observed at any redshift \cite[e.g.][]{Noeske2007a, Elbaz2007, Daddi2007a}, corresponds to a well defined sequence of disk galaxies, while spheroidal systems tend to live below the MS. In light of this finding, the SFR--density relation can be seen as an alternative way to study the morphology--density relation.

A galaxy's SFR is on average anti-correlated with the galaxy density in the local Universe  \citep{Lewis2002, Gomez2003, Kauffmann_etal_2004}. In fact, highly-star-forming galaxies are mostly found in low-density environments, while the cores of massive clusters are full of massive, early-type galaxies dominated by old stellar populations. However, the way this relation evolves with redshift is still a matter of debate. 

It has been argued that as we approach the epoch at which early-type galaxies form the bulk of their stars at $z \gtrsim 1.5$ \cite[e.g.][]{Rettura2010}, the SFR--density should progressively reverse, such that high-density regions host highly star-forming galaxies at earlier cosmic time. 
\cite{Elbaz2007} and \cite{Cooper2008} observe this reversal already at $z\sim1$ in the GOODS field and the DEEP2 Galaxy Redshift Survey, respectively. Using {\it{Herschel}} PACS (Photodetecting Array Camera and Spectrometer, \citealt{poglitsch10}) data, \cite{popesso11}  detect the reversal only for high-mass galaxies. According to the authors, this is due to high-mass galaxies being more likely to host Active Galactic Nuclei (AGN). Since AGN exhibit a slightly higher SFR with respect to galaxies of the same stellar mass \citep{Santini2012}, AGN hosts tend to be star-forming (see also \citealt{Rosario2013}). On the other hand, \citet{Feruglio2010}, \cite{Ideue2009, Ideue2012} and \cite{Tanaka2012} find no reversal in the COSMOS field, arguing that the reversal, if any, must occur at $z\sim 2$.

The aforementioned studies, use different SFR indicators.
\cite{Cooper2008} and \cite{Muzzin2012} convert the [OII] emission line flux into a SFR, while \cite{Elbaz2007}, \cite{Feruglio2010}, and \cite{Tran2010} use {\it Spitzer}  MIPS 24~\um data to measure the SF activity of their galaxy sample. In addition, \cite{Elbaz2007} complement the estimates of SFR derived from the 24~\um flux with those from ultra-violet (UV) emission. All of these estimators can be heavily affected by dust extinction uncertainties, by AGN contamination, and/or by metallicity (e.g. \citealt{Kewley2004}). These problems can be overcome by measuring the SFR from the far-infrared (IR) luminosity, as done in \cite{popesso11}. Indeed, {\it{Herschel}} PACS data cover the wavelength range at which the bulk of the UV light is re-emitted by dust, at least up to $z\sim1.5$ \citep{Elbaz2011}. This enables an accurate estimate of the SFR and avoids possible contamination by AGN emission, more common in the mid-IR spectral range \citep{Netzer2007}. 

Also the definition of the environment estimated via the local galaxy density is somewhat arbitrary. Indeed, several works measure the distance to the Nth nearest neighbour \cite[e.g.][]{Cooper2008}. This method is strongly dependent on N: small values probe high-density regions better though they smooth the low-density ones,  while high values of N could wash out the information on over-densities when the number of galaxies in a given halo is less than N \citep{Cooper2005, Muldrew2012, Woo2012}.  Other authors measure the density of neighbours within a fixed co-moving volume centred on each galaxy \cite[e.g.][]{Elbaz2007, popesso11}. 

All of these methods rely on the assumption that the local number density of galaxies is a good representation of the environment. However, if the environment is defined as the halo mass of the parent halo to which the galaxy belongs, this is not necessarily the case. Indeed, a filament (interconnecting ``nodes'' of the same large scale structure), the outskirts of a massive galaxy cluster, and the core of a galaxy group could exhibit the same galaxy density, even being sites of quite different physical processes (on multiple scales these environments can be separated, see e.g. \citealt{Wilman2010}).

Further complication is added by the interplay of mass and density. According to \cite{Kauffmann_etal_2004}, mass and galaxy density are coupled, with the high-mass galaxies segregated in the densest environments. This relation was already in place at $z \sim 1$ \citep{Scodeggio2009, Bolzonella2010}. Therefore, the evidence for a clear SFR--density trend could be due to a different contribution of massive and less massive galaxies favouring different density regimes.

In order to shed light on the relation between SFR, density, and halo mass, we take advantage of the combination of the deepest available {\it{Spitzer}} and {\it{Herschel}} surveys of the Extended Chandra Deep Field South (ECDFS) and the GOODS-South and -North fields (GOODS-S and GOODS-N, respectively), observed in the PACS Evolutionary Probe (PEP, \citealt{Lutz2010}) and GOODS-{\it Herschel}  \citep{Elbaz2011} surveys. The combined GOODS data from these two surveys are described in \cite{magnelli2013}. In this work we use a spectroscopic selected sample as already done in \citet[Z13 hereafter]{Ziparo2013}.

We first study the SFR--density relation up to $z\sim1.6$ in its standard definition, by estimating the local galaxy density parameter. In the second part of the paper, we propose an alternative definition of the SFR--environment relation: we distinguish between galaxy group members, ``filament-like'' environments and galaxies that are isolated or more likely associated with lower mass haloes. For this analysis, we use the galaxy group sample studied in Z13. In addition, we try to break the mass--density (environment) degeneracy, by studying the location of group galaxies in the SFR--$\rm M_\star$ ($\rm M_\star$) plane as a function of environment up to $z\sim 1.6$.

This paper is organized as follows: in Section~\ref{dataset} we briefly describe our dataset and analysis. In Sec.~\ref{sfr_density} we present our results and  we discuss them in Sec.~\ref{discussion}. Eventually, we draw our conclusions in Sec.~\ref{summary_conclusions}. Throughout our analysis we adopt a \cite{Chabrier2003} initial mass function (IMF) and the following cosmological parameters: $\mathrm{H_0=70} \mathrm{\ km\ s^{-1}\ Mpc^{-1}}$, $\Omega_\mathrm{M}=0.3$ and $\Omega_\mathrm{\Lambda}=0.7$.

%----------------------------------------------------------------------------------------

\section{The Dataset}
\label{dataset}
%----------------------------------------------------------------------------------------
%
In Z13 we create a clean IRAC (Infrared Array Camera, \citealt{Fazio2004}) 3.6 $\mu$m selected galaxy sample in the ECDFS and GOODS fields. This sample includes only galaxies with a spectroscopic redshift and is drawn from the galaxy catalogues of  \cite{Cardamoneetal2010}, \cite{grazian06}, and \cite{Wuyts2011}, in the ECDFS, the GOODS-S and the GOODS-N field, respectively. The group sample studied in Z13 also includes the X-ray groups identified in the COSMOS field by \cite{Finoguenov2007, George2011}, and \cite{George2012}, and employs the group membership defined by \cite{popesso2012}. 
However, given the rather low spectroscopic completeness in the COSMOS field (40\% in the $\rm M_\star$ range of interest, see Z13 for a complete discussion), this region is not included in our current analysis. Indeed, it is not possible to reliably estimate the local galaxy density parameter on the basis of the pure spectroscopic data.  The use of both spectroscopic and photometric redshifts, as done in \cite{Kovac2010}, is preferable in the COSMOS field, where the sampling rate is spatially very inhomogeneous. Thus, since the ECDFS and the GOODS fields show an extremely high spectroscopic completeness (60-80\% in $\rm M_\star$), we prefer to restrict our analysis to these regions.

We measure the SFR by using the deepest available {\it{Spitzer}} Multiband Imaging Photometer (MIPS) 24 $\mu$m data combined with the deepest {\it{Herschel}} PACS  100 and 160 $\mu$m data. 
In order to overcome any blending issue, the {\it Spitzer}  and {\it Herschel} flux densities are derived with a point-spread-function-fitting analysis guided by the position of sources detected in deep IRAC images \cite[see][]{magnelli11,magnelli2013}. This method solves a large part of the blending issues encountered (see results of dedicated Monte-Carlo simulations in \citealt{magnelli2013}) and provides a straightforward association between IRAC, MIPS and PACS sources. Furthermore, even if in high density regions the prior PSF-fitting method does not solve all blending issues, it should still provide reliable estimates of the total infrared fluxes of these clustered regions and, thus, of their total SFR activity.

The SFR is estimated with the use of the IR templates of \cite{Elbaz2011}. For sources undetected in PACS and with only MIPS detections, we use the ``Main Sequence'' template, which turns out to provide the most accurate estimate of the SFR from mid-IR data. In order to complement the SFR derived from IR data (available for the bulk of the star-forming population) with the SFR of the low-star-forming or rather inactive galaxies (i.e. undetected in the mid- and far-IR surveys), we measure the SFR via multi-wavelength SED fitting by using {\it Le PHARE\footnote{http://www.cfht.hawaii.edu/~arnouts/LEPHARE/cfht\_lephare/ lephare.html}} (PHotometric Analysis for Redshift Estimations; \citealt{arnoutsetal2001, ilbertetal2006}) and the  \cite{bruzual_charlot2003} library. For this purpose we use the aforementioned multi-wavelength photometric catalogues \citep{Cardamoneetal2010, grazian06, Wuyts2011}. 

Z13 provide a careful calibration of the SFR derived via SED fitting with respect to the more reliable SFR derived from IR data. We find consistent estimates of SFR though the scatter is quite large, ranging from 0.5 to 0.7 dex depending on the redshift range.

The SED fitting technique is also useful for estimating stellar masses. The comparison of our estimates with those derived from the same catalogues via different methods and/or templates shows that we can accurately estimate $\rm M_\star$ within a factor of two (see Z13 for more details). 

The spectroscopic data used for the construction of the density field and the dynamical analysis of the galaxy group sample are taken from a collection of publicly available high-quality spectroscopic redshifts in the ECDFS (\citealt{Cardamoneetal2010, Cooper2011, Silverman2010, balestraetal2010, Popesso2009}, see Z13 for further details about the combination of the different catalogues). The spectroscopic catalogue of the GOODS-N field is taken from \cite{Barger2008}.

\subsection{The galaxy group sample and their members}

\begin{figure}
\centering
   \includegraphics[width=\hsize]{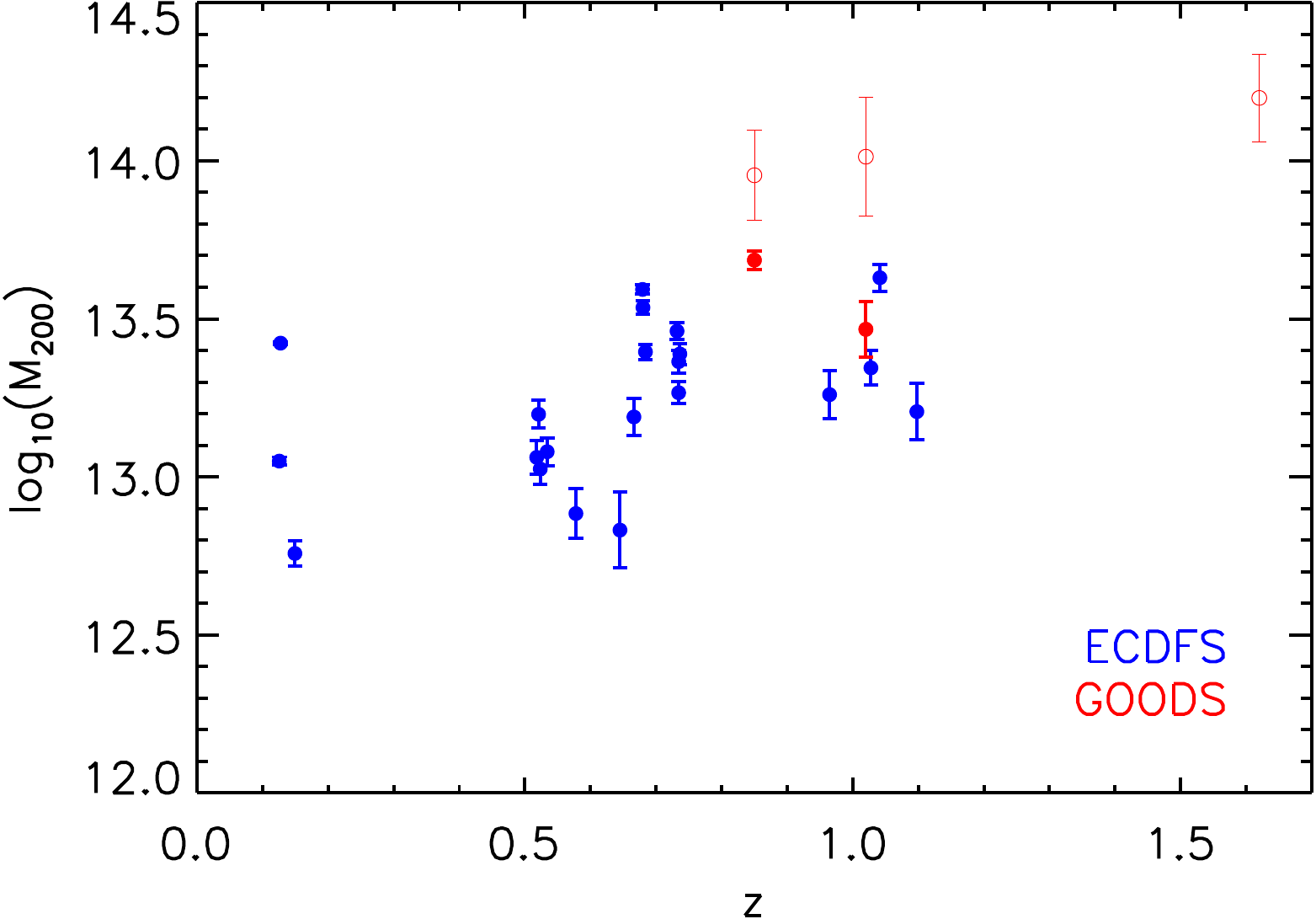}
  \caption{$\rm M_{200}$ as a function of redshift for all groups considered in our sample. Filled circles represent the X-ray mass estimates, while empty circles show the dynamical mass estimates. We highlight in blue the ECDFS sample and in red the GOODS groups.
}
  \label{fig:m200_z}
\end{figure}

All the blank fields considered in our analysis are observed extensively in the X-ray with $Chandra$ and XMM-$Newton$. The X-ray data reduction and the creation of the X-ray group catalogues are explained in detail in \cite{Finoguenov2009} and in Finoguenov et al. (in prep.). 
As explained in Z13, we select a sub-sample of X-ray selected groups with clear optical (spectroscopic) identification (we do not include groups with more than one redshift peak of similar strength along the line of sight), without close companions that might affect the membership determination, and with at least 10 members, to reliably estimate the velocity dispersion and the membership. This selection leads to a sample of 22 X-ray detected groups in the ECDFS and 2 groups in the GOODS-N field.  We also consider a large scale structure spectroscopically confirmed at $z\sim 1.6$ by  \cite{Kurk2009}.

Fig.~\ref{fig:m200_z} shows the group mass\footnote{$\mathrm{M_\Delta}$ (where $\mathrm{\Delta=500,200}$) is defined as $\mathrm{M_\Delta=(4 \pi/3) \Delta \rho_c R_{\Delta}^3}$, where $\mathrm{R_\Delta}$  is the radius at which the density of a cluster is equal to $\Delta$ times the critical density of the Universe ($\mathrm{\rho_c}$)} estimates as a function of redshift for the ECDFS (in blue) and GOODS (in red) fields, respectively. We also show the dynamical mass estimates for the groups in the GOODS fields from \cite{popesso2012}.  
The dynamical analysis of each structure is based on spectroscopic data. For details on the dynamical analysis and group membership, see \cite{popesso2012} and Z13.

In order to follow the evolution of the relation between SFR and environment, we divide our galaxy sample into four redshift bins, $0<z\leq 0.4$,  $0.4<z\leq 0.8$, $0.8<z\leq 1.2$, and $1.2<z\leq 1.7$, according to the redshift distribution of our group sample. We note that the last redshift bin is populated only by the structure at $z\sim 1.6$ \citep{Kurk2009}. This is a likely super-group or a cluster in formation as suggested by the X-ray emission from different extended sources in the structure (Finoguenov et al. in prep.). When we analyse the SFR--environment relation by distinguishing group members from systems in other environments, we consider, in each redshift bin, all group galaxies together as members of a composite group. 
This is done to increase the statistics of group galaxies which otherwise would be too low when considering individual systems.

To limit the selection effects and at the same time to control the different levels of spectroscopic completeness in different redshift bins (see e.g. fig.~5 in Z13), we apply a common stellar mass cut at $\rm M_\star =10^{10.3}~M_{\odot}$. This mass cut corresponds to an IRAC 3.6 $\mu$m apparent magnitude brighter than the 5$\sigma$ detection limit in each considered field up to $z\sim 1.7$, enabling a high spectroscopic completeness. Moreover, the considered mass range is still dominated by sources with MIPS and/or PACS detections, in other words with robust SFR estimates.
The uncertainties due to the spectroscopic incompleteness of our galaxy sample is evaluated with dedicated Monte Carlo simulations based on the mock catalogues of \cite{KW_millennium2007} drawn from the Millennium simulation \citep{Springel2005}.

\subsection{The local galaxy density}
\label{sec:galaxy_density}

The key ingredient for building a reliable density field is very high and spatially uniform spectroscopic coverage. This is reached in the ECDFS (see \citealt{Cooper2011}) and in the GOODS fields (see \citealt{popesso11, Elbaz2007}), for which we reconstruct the density around each galaxy up to $z\sim 1.7$.

We use a method similar to \cite{popesso11} to compute the projected local galaxy density, $\Sigma$, around each spectroscopically-confirmed galaxy with $\rm M_\star > 10^{10.3} M_{\odot}$. We count all galaxies located inside a cylinder of radius $\rm 0.75~Mpc$, within a fixed velocity interval around each galaxy of $\rm \Delta {v}=3000~\rm{km}\,{s}^{-1}$, about ten times the typical velocity dispersion of galaxy groups ($\sigma_v \sim 300-500$ $\rm{km}\,{s}^{-1}$), and above a redshift dependent mass limit  ($\rm M_{cut}(z)$). Given the spectroscopic completeness as a function of $\rm M_\star$ in the four redshift bins considered in our analysis (see Z13), we choose as a cut the $\rm M_\star$ value where the 40-50\% completeness limit is reached in each redshift bin: $\rm M_\star/M_{\odot} = 10^{9}$ at $0 < z < 0.4$, $\rm M_\star/M_{\odot} = 10^{9.5}$ at $0.4 < z < 0.8$, $\rm M_\star/M_{\odot}=10^{10} $ at $0.8 < z < 1.2$ and  $\rm M_\star/M_{\odot} = 10^{10.3}$ at $1.2 < z < 1.7$.  
This does not lead to a different density field definition as a function of redshift bin, but only to a more robust density estimate in the bins where the spectroscopic completeness is still very high at low masses. Indeed, only the absolute value of the density parameter changes, but the relative difference between high and low density regions is kept the same with respect to the choice of a fixed $\rm M_\star=10^{10.3}M_{\odot}$ at any redshift. 

The density field obtained with the chosen $\rm M_{cut}(z)$, rather than that at a fixed mass cut of  $\rm M_\star=10^{10.3}M_{\odot}$, allows us to distinguish between galaxies residing in dark matter haloes of different masses. However, the density fields obtained with a lower mass cut show,  as expected, higher values and a slightly higher accuracy in distinguishing between galaxies located in parent haloes of different masses.  We estimate that, on average, the projected density obtained with a mass cut of $\rm M_\star/M_{\odot} = 10^{9}$ at $0 < z < 0.4$ is a factor of 7 higher than that at $\rm M_\star/M_{\odot} = 10^{10.3}$. At $0.4 < z < 0.8$, a mass cut of  $\rm M_\star/M_{\odot} = 10^{9.5}$  leads to a density   a factor of 5 higher than the cut at lower $\rm M_\star$, while at $0.8<z<1.2$ the density with a mass cut of $\rm M_\star/M_{\odot}=10^{10} $ is a factor of 2.5 higher. These calibrations are discussed in depth in a dedicated forthcoming paper (Popesso et al., in prep.).

A more physical definition of the density field would require a mass cut which takes into account the evolution of the characteristic magnitude of the stellar mass function. In our case this would translate to selecting only galaxies at masses larger than $\rm M^*$ at any redshift \citep{ilbertetal2010}, given the restriction imposed by the completeness level of the galaxy sample in the higher redshift bins.  This would imply that at lower redshifts we would select only galaxies with $\rm M_\star >10^{11}\,M_{\odot}$, limiting in a significant way the statistics for defining the density field. Thus, distinguishing between galaxies residing in parent haloes of different masses would be inefficient. 
A simple exercise on the mock catalogues of \cite{KW_millennium2007} reveals that this density definition would be able to distinguish only isolated galaxies from galaxies in the core of massive clusters and it would provide the same density for galaxies in haloes with masses ranging from $10^{12}$ to $\rm 10^{14.5}\,M_{\odot}$. 

In order to consider the effect of spectroscopic incompleteness, we correct the density $\Sigma$ by accounting for the possibly missing galaxies. We consider, for each galaxy, the cylinder along the line of sight, with radius of 0.75~Mpc at the redshift of the considered source, and with redshift limits $z_{min}$ and $z_{max}$ equal to the limits of the redshift bin to which the source belongs. The spectroscopic completeness is given by the number of sources with spectroscopic redshifts divided by the total number of galaxies, considering only sources with $\rm M_\star>\rm M_{cut}(z)$. Since the redshift bins are more than 10 times larger than the error on the photometric redshift, this uncertainty is only marginally affecting our completeness estimate.
We correct for incompleteness by dividing $\Sigma$ by this ratio. 

\begin{figure}
\centering
   \includegraphics[width=\hsize]{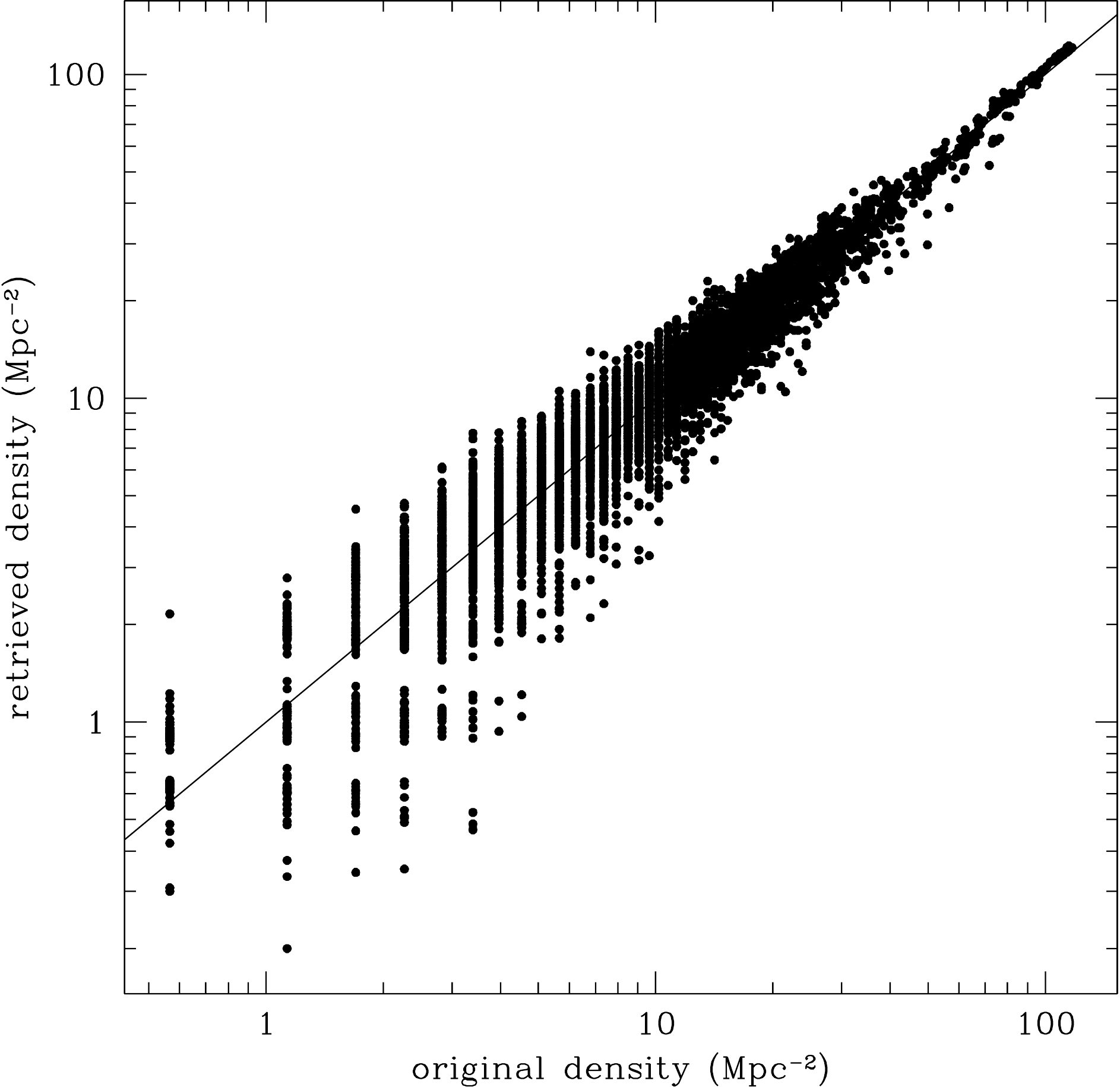}
  \caption{Comparison between the original local density estimated in the \citet{KW_millennium2007} mock catalogues and the density retrieved in the randomly extracted catalogue following our method. The solid line shows the 1-to-1 relation.
}
  \label{comp}
\end{figure}

In order to test the reliability of our density estimate, we measure, with the same method, the density field in 100 randomly-extracted catalogues from the mock catalogues of \cite{KW_millennium2007}. We compare the density obtained in this way with that measured in the parent light-cone mock catalogues, free of selection biases.  In order to simulate the photometric redshift uncertainty, before estimating the incompleteness correction, we assign a random error in the range $\rm -{\Delta}z < {\delta}z_{phot} < {\Delta}z$ to the redshift of the parent mock catalogue galaxies, where ${\Delta}z$ is the photometric redshift error provided in the photometric catalogues. We find a very good agreement between the original density estimated in the \cite{KW_millennium2007} mock catalogues and the local density retrieved with our method (Fig.~\ref{comp}). We also use this approach for estimating the error per density bin as the dispersion of $\rm \Sigma_{original}-\Sigma_{retrieved}$.

\begin{figure}
\centering
   \includegraphics[width=\hsize,angle=180]{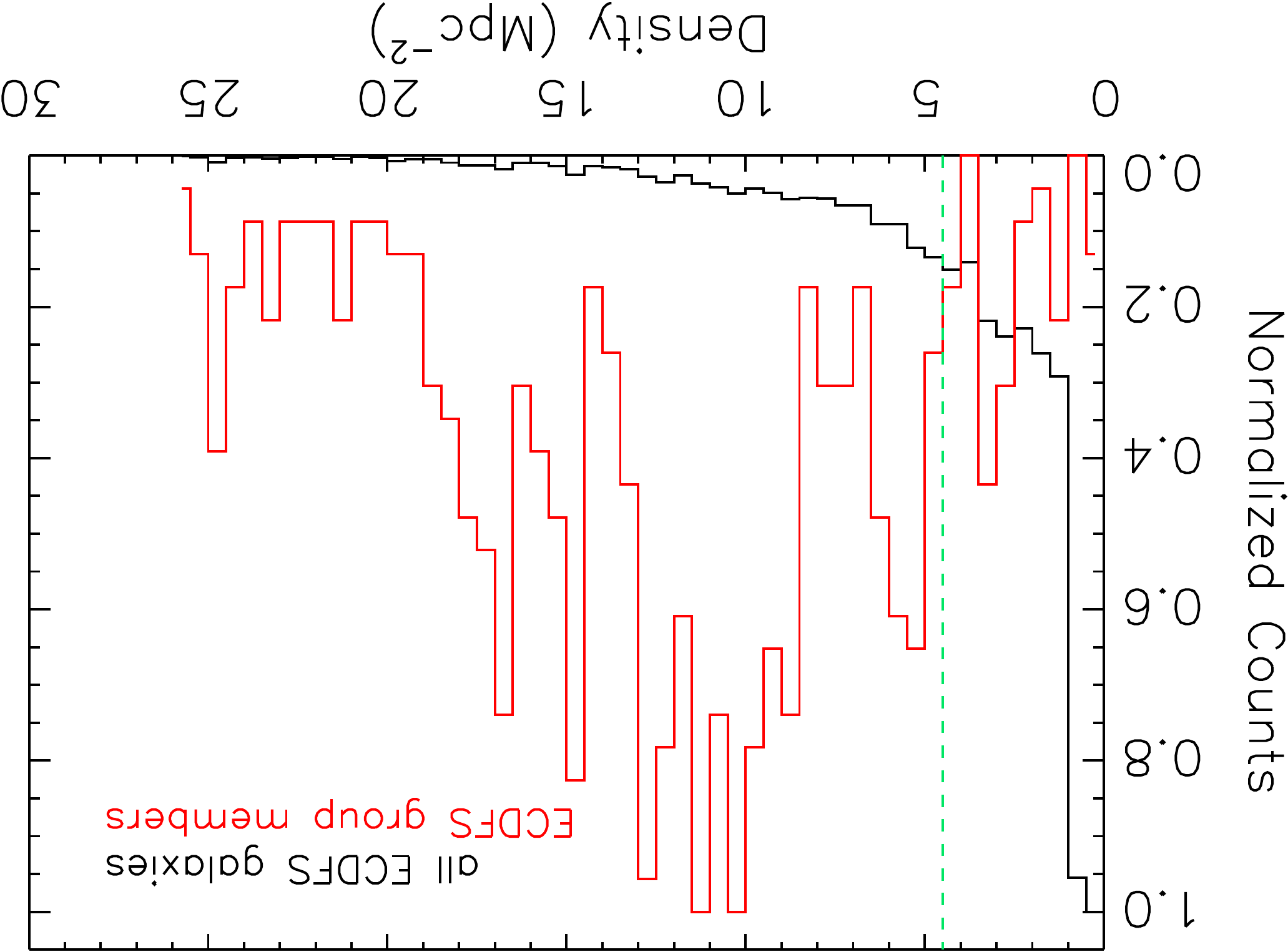}
  \caption{In black: density distribution around each galaxy with a spectroscopic redshift in ECDFS. The red histogram shows the density of group members. The green dashed line at ${\rm 4.5\ galaxies\ Mpc^{-2}}$  nicely separates the group-dominant regime from the field-dominant regime. Indeed, 75\% of field galaxies are found at densities below this threshold and 92\% of group galaxies above that.
}
  \label{density}
\end{figure}

Our density definition takes advantage of the high level of mass segregation observed in the local Universe \citep{Kauffmann_etal_2004} and at least up to $z\sim 1$. Since we estimate the density of rather massive galaxies around each system, our density estimator should be able to better distinguish between high-density regions, generally dominated by massive galaxies, from low density regions, more populated by low-mass systems.  Indeed, Fig.~\ref{density} shows that our method is able to nicely isolate galaxies identified as group spectroscopic members (red histogram) from isolated galaxies (the peak below $\rm \Sigma \sim 3-4\,Mpc^{-2}$). A similar figure is shown in \citet[their fig.~11]{Cooper2011} based on the third-nearest neighbour density estimator. The comparison of the two figures shows that our density estimator is more efficient in distinguishing isolated systems from galaxy group members. In fact, although groups occupy the highest density bins in \citet[their fig.~11]{Cooper2011}, they are not clearly isolated from field galaxies as in our case.

%----------------------------------------------------------------------------------------

\section{Results}

\label{sfr_density}
%----------------------------------------------------------------------------------------

We first build the SFR--density relation by studying the statistical correlation between the SFR and density parameters, as usually done in the literature. This lets us compare our results with previous works. As a second approach we use a dynamical definition of environment by differentiating among massive bound structures, less-massive bound or unbound structures and relatively isolated galaxies. We follow the evolution of the relation in both cases up to $z\approx 1.6$ and we test and compare our results with the predictions of simulations.

\subsection{The ``environmental'' approach}
\label{SFR_density_environmental}

\begin{figure}
\begin{center}
\includegraphics[width=\hsize,angle=180]{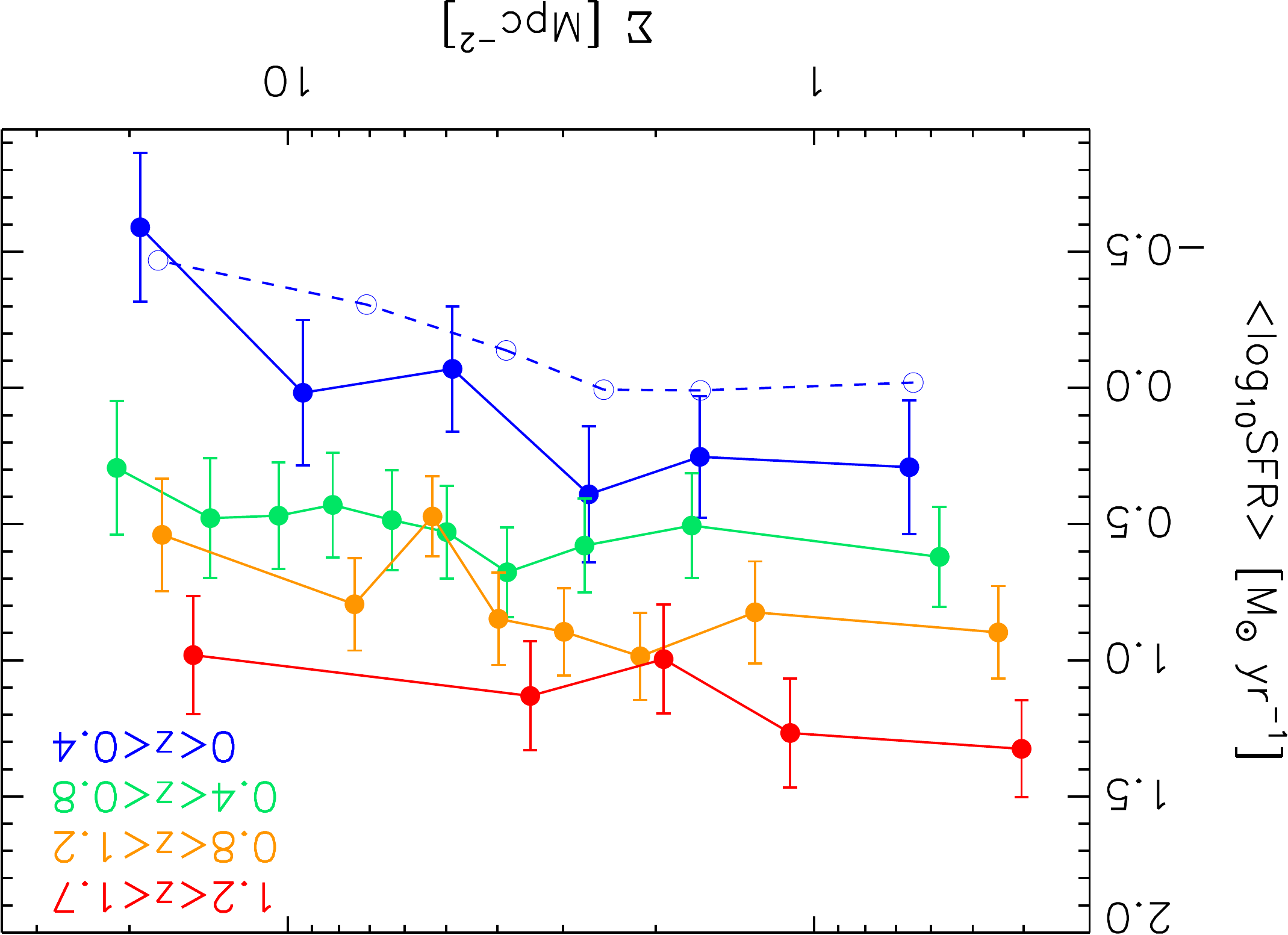}
\end{center}
\caption{SFR--density relation for galaxies with $\rm M_\star>10^{10.3}\,M_\odot$ in different redshift bins (solid lines). The dashed line represents the SFR--density relation at $0 < z < 0.4$ for all galaxies with $\rm M_\star> 10^9\,M_{\odot}$. Errors are derived using the mock catalogues of \citet{KW_millennium2007}, as explained in the text.}
\label{SFR_density_relation}
\end{figure}

Fig.~\ref{SFR_density_relation} shows the SFR--density relation for all galaxies with  $\rm M_\star >10^{10.3}\,M_{\odot}$ in four redshift bins.  
The errors in Fig.~\ref{SFR_density_relation} are derived from our error analysis using the mock catalogues of \cite{KW_millennium2007}, as explained in Sec.~\ref{mock_errors}. 
We find a significant anti-correlation up to $z\sim 0.8$, confirmed by the Spearman test at  $3\sigma$ confidence level. At $0.8< z< 1.2$ we find an anti-correlation but with lower significance ($2.3\sigma$). In the highest redshift bin, comprising the \cite{Kurk2009}  large scale structure, we do not find any significant anti-correlation ($< 2\sigma$ significance level). Thus, we can exclude with high confidence level (from the Spearman test) any positive correlation in the last two redshift bins as claimed in previous works \cite[e.g.][]{Elbaz2007, Cooper2008}. We only observe a progressive flattening towards higher redshifts, but no reversal of the relation.

The shapes of the relations shown in Fig.~\ref{SFR_density_relation} are noisy and not even linear in log-log space. Thus, it can not be easily fit by a simple fitting function. In order to quantify the steepness of the relation, we simply estimate the ratio between the mean SFR at densities below and above the median local galaxy density $\Sigma$. Below $z\sim 1.2$, where we see an anti-correlation, although with different significances depending on the redshift bin, the mean SFR in low density regions spans a range of 1.4-2.1 times the mean SFR in high density regions. In the highest redshift bin we do not observe a significant difference between the SFR in low and high density regions. 

What is the role of group galaxies in shaping the relations? In order to check this, we remove from the sample all galaxies dynamically associated with either extended X-ray emitting sources or the structure at $z\sim 1.6$. We also remove all galaxies associated with extended X-ray emitting sources not included in the final group sample. 
In the two lowest redshift bins, the significance of the anti-correlation decreases much below the $3\sigma$ level. In the highest two redshift bins we exclude both a reversal and any sign of anti-correlation. This clearly shows the dominant role of group environments in shaping the SFR--density anti-correlation observed in the local Universe and at intermediate redshift.

\begin{figure}
\centering
\includegraphics[width=\hsize,angle=180]{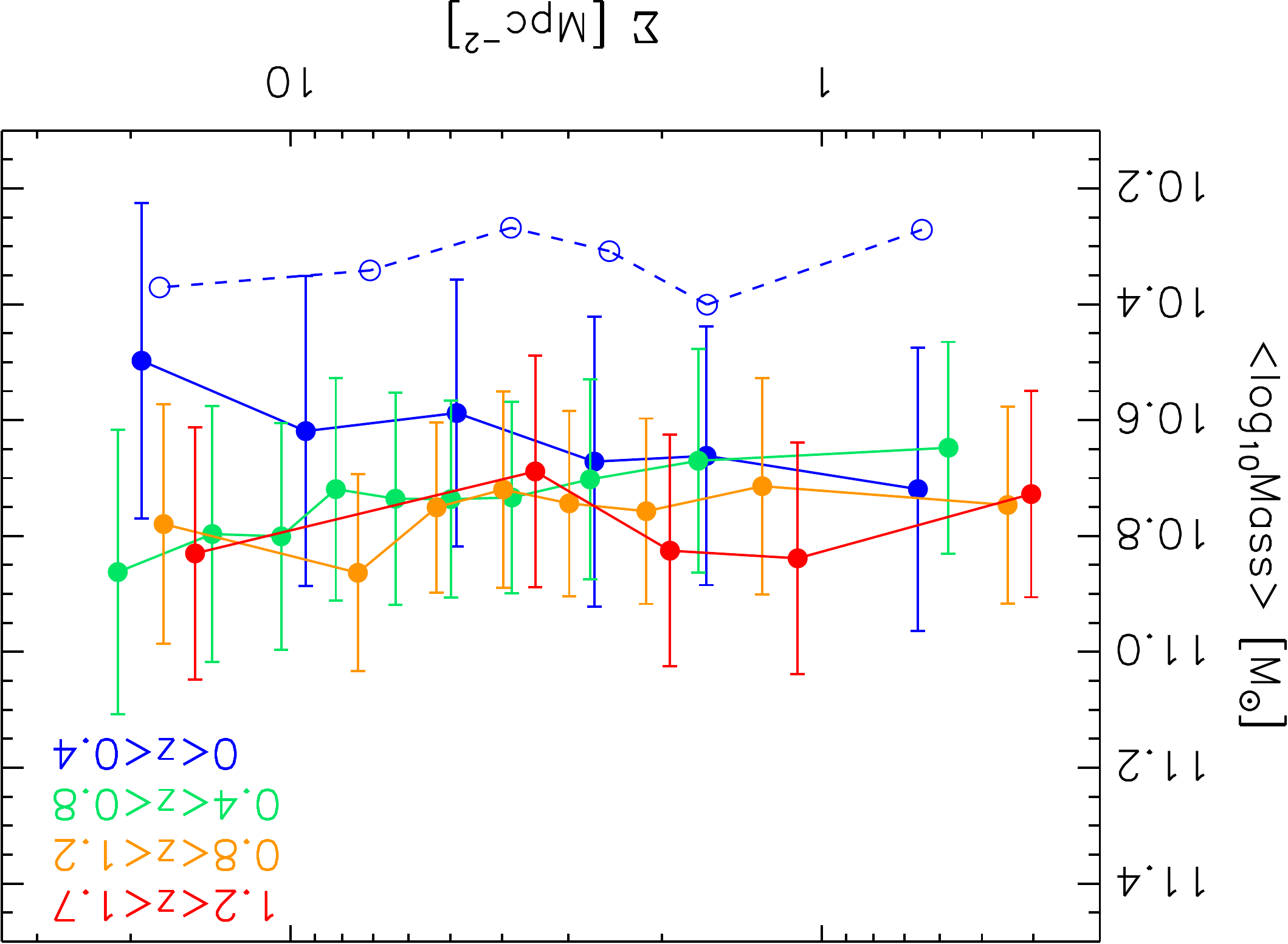}
\caption{Stellar mass---density relation for all the galaxies with $\rm M_\star>10^{10.3}\,M_\odot$ in different redshift bins (solid lines).  Errors are derived using the mock catalogues of \citet{KW_millennium2007}, as explained in the text. The dashed line represents the $\rm M_\star$--density relation at $0 < z < 0.4$ for all galaxies with $\rm M_\star> 10^9\,M_{\odot}$. 
The normalisation of the dashed line is artificially increased to higher value to make it close to the blue solid line and make the comparison easier.
}
\label{mass_density}
\end{figure}

In principle, a prominent mass segregation together with a high fraction of low-star-forming galaxies (typical of the group and cluster environment) could easily lead to the SFR--density anti-correlation observed at low and intermediate redshift. Fig.~\ref{mass_density} shows the $\rm M_\star$--density relation for the same sample of galaxies in the four redshift bins. 
We do not measure a strong mass segregation in the galaxy sample used for the SFR--density relation analysis. 
The first redshift bin exhibits a slightly different behaviour with respect to the relation in the other redshift bins. 
However, given the large errors (for their computation see Sec.~\ref{mock_errors}), we can not draw a definitive conclusion on the $\rm M_\star$--density trend. 
The Spearman test confirms only a mild level of mass segregation at $0.4 < z < 0.8$. 
Thus, the SFR--density anti-correlation observed in the first and second redshift bins is not caused by mass segregation. 

In the local Universe, mass segregation is observed with high significance by \cite{Kauffmann_etal_2004} on the basis of a large sample of Sloan Digital Sky Survey (SDSS, \citealt{York2000}) galaxies. Are our results at odds with previous findings? The main difference with respect to 
\cite{Kauffmann_etal_2004} is the mass cut applied to our sample. Indeed, for spectroscopic completeness issues, we are considering only massive galaxies, i.e with $\rm M_\star>10^{10.3}\,M_{\odot}$. The dashed blue line in Fig.~\ref{mass_density} shows the mass--density relation obtained after applying a mass cut of $\rm 10^9\,M_{\odot}$ in the lowest redshift bin. This analysis is possible without strong biases only at low redshift where the spectroscopic completeness is rather high even at low stellar masses (see Z13). The Spearman test reveals a mild positive correlation. The absence of a stronger correlation, as found e.g. in \cite{Kauffmann_etal_2004}, could be due to the lack of many massive spectroscopically-detected galaxies at low redshift (see fig.~5 of Z13), since in ECDFS this type of galaxies was targeted for spectroscopy only at high redshift \cite[e.g.][]{Popesso2009}.  
We point out that we observe an even more significant (according to the Spearman test) SFR--density anti-correlation (blue dashed line in Fig.~\ref{SFR_density_relation}) in the lowest redshift bin after applying the lower mass cut. This probably indicates that, in a broader mass regime, mass segregation enhances the significance of the SFR--density relation in the parent sample.

\subsubsection{How robust is our analysis of the SFR--density relation?}
\label{mock_errors}
In order to take into account all possible biases inherent in our spectroscopic selection, we study the SFR--density relation in a simulated Universe. 
We analyse the SFR--density relation obtained using the \cite{KW_millennium2007} mock catalogues (5 different light cones) by applying our definition of local galaxy density. We observe an anti-correlation in all redshift bins ($5\sigma$ significance). Thus, our results (Fig.~\ref{SFR_density_relation}) are at least qualitatively in agreement with the prediction of \cite{KW_millennium2007}, except in the highest redshift bin.  However, we point out that the SFR--density relations measured using the mock catalogues are observed in an area of sky that is a magnitude larger than the ECDFS and the GOODS-N regions. Thus, the mock catalogues sample a much broader range of densities due to the presence of massive clusters, while our dataset comprises only groups. 

In order to check the effect of cosmic variance when using a rather small area, we estimate the SFR--density relation in 1000 different regions of the \cite{KW_millennium2007} light-cones with areas similar to the sum of the ECDFS and GOODS-N areas. After running a Spearman test, we detect an anti-correlation with at least $3\sigma$ significance in all regions in the two redshifts bins below $z\sim0.8$. At $0.8<z <1.2$ we measure an anti-correlation in 98\% of the cases and at higher redshift in 70\% of the cases.  The non-correlation in the observed $1.2<z<1.7$ redshift bin could be due to the low probability of finding massive large scale structures in such a small area and at high redshift in the $\Lambda$CDM cosmology. It could also be due to larger errors on environment washing out the signal (see e.g. \citealt{Cooper2010}).Thus, cosmic variance could considerably affect the significance of an anti-correlation.

In order to simulate the spectroscopic completeness in the ECDFS and GOODS regions, we randomly extract a sub-sample of galaxies from the Millennium mock catalogues mimicking the spectroscopic completeness of the observed data. We extract randomly a percentage of galaxies consistent with the spectroscopic completeness in one of the available photometric bands and for each magnitude bin of our galaxy sample. 
This procedure randomly extracts 100 different catalogues that nicely reproduce the selection function of our sample (see Z13 for further details). We point out that while the galaxy mock catalogues of the Millennium simulation provide a rather good representation of the local Universe, at higher redshifts ($z > 1$) they fail to reproduce the correct distribution of star-forming galaxies in the SFR-$\rm M_\star$ plane. Indeed, \cite{Elbaz2007} find that at $0.8 < z < 1.2$ the galaxy SFR is under-estimated, on average, by a factor of two,  at fixed $\rm M_\star$, with respect to the observed values. By performing the same exercise with our dataset, we find that this under-estimate ranges by factors  of 2.5 to 3 at $1.2 < z < 1.7$. 
However, this does not represent a problem with our approach. Indeed, the aim of this analysis is to understand what is the bias introduced by a selection function similar ``in relative terms'' to the spectroscopic selection function of our dataset. Thus, for our needs, it is sufficient that the randomly extracted mock catalogues reproduce the same bias in selecting, on average, the same percentage of most star-forming and most massive galaxies of the parent sample, as shown in Z13.
The bias of our analysis is estimated by comparing the results obtained with and without our galaxy sampling. 
Since the under-estimate of the SFR or the $\rm M_\star$ of high redshift galaxies is common to both, biased and unbiased, samples, it does not affect the result of this comparative analysis. We also stress that the aim of this analysis is not to provide correction factors for our observational results but a way to interpret our results taking into account possible biases introduced by the spectroscopic selection function.

In order to account for both the spectroscopic completeness and the cosmic variance, we repeat the exercise performed with the complete \cite{KW_millennium2007} mock catalogues by extracting 1000 different regions with areas similar to the sum of the ECDFS and GOODS-N areas in the incomplete mock catalogues. We estimate the SFR--density relation in each region as done on the real dataset. The probability of non-correlation increases to $\sim$5\% in lower redshift bins,  12\% at $0.8<z <1.2$  and 45\% at $1.2 < z < 1.7$. This suggests that small areas (thus cosmic variance), in addition to spectroscopic incompleteness, could hide a possible anti-correlation in the highest redshift bin or reduce the significance of the anti-correlation in the lower redshift bins.

This last exercise allows us to quantify the possible bias in the estimate of the SFR--density relation due to our spectroscopic selection. We measure the mean SFR by using the same binning in density in the incomplete and in the original complete catalogues. In this way, we can compute the residual $\rm \Delta SFR (\Sigma)=\langle SFR_{observed}(\Sigma)\rangle/\langle SFR_{true}(\Sigma)\rangle$, where $\rm \langle SFR_{observed}\rangle$ is the mean SFR estimated in the incomplete catalogue at the given density bin, and $\rm \langle SFR_{true}\rangle$ is the mean SFR estimated in the complete \cite{KW_millennium2007} mock catalogues at the same density bin. We estimate $\rm \Delta SFR (\Sigma)$ in 1000 sky regions, extracted from 5 light cones, with the area similar to the sum of the ECDFS and GOODS-N area, as explained above. 
We estimate the mean and the dispersion of the $\rm \Delta SFR (\Sigma)$ distribution in each density bin. The mean indicates whether there is any bias in the spectroscopic selection that leads to an over- or under-estimate of the mean SFR per density bin. The dispersion provides the error on the mean SFR per bin.

\begin{figure}
\centering
\includegraphics[width=\hsize]{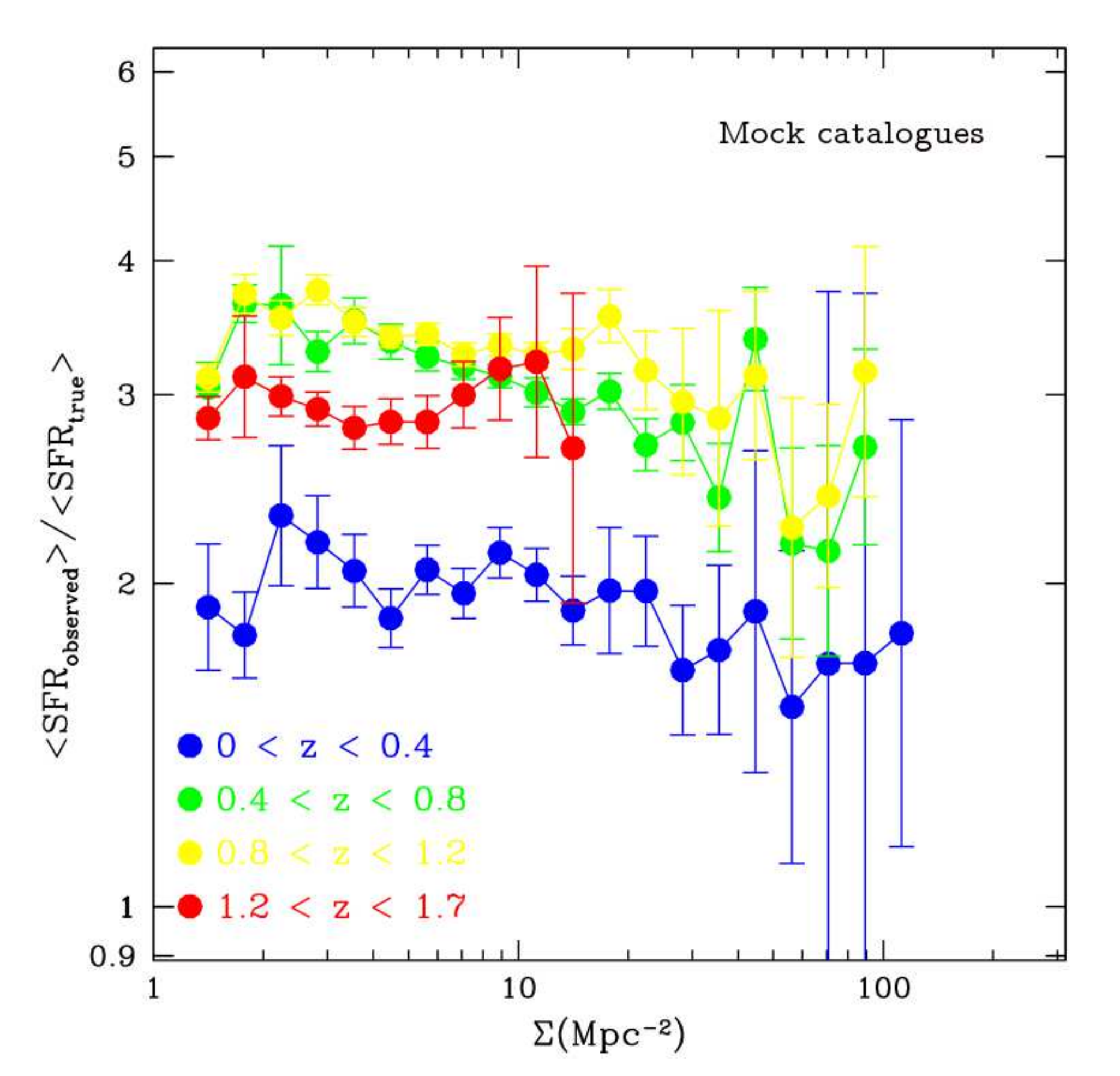}

\caption{Ratio between the SFR derived from the incomplete mock catalogues ($\rm \langle SFR_{observed}\rangle $) and the SFR from the complete ones ($\rm \langle SFR_{true}\rangle $) versus density for all four redshift bins used in this work. We do not find any bias in the slope of the SFR--density relation of Fig.~\ref{SFR_density_relation} as confirmed by the Spearman test.}
\label{SFR_density_nobias}
\end{figure}

As shown in Fig.~\ref{SFR_density_nobias}, the $\rm \langle SFR \rangle$ derived from the incomplete mock catalogues is on average a factor of 2-4 (depending on the redshift bin) larger than the ``true'' one obtained from the complete \cite{KW_millennium2007} mock catalogues. Thus, the incompleteness leads to a large over-estimate of the mean SFR in each density bin. This is easily understandable, since the simulated spectroscopic selection favours highly-star-forming galaxies (see Z13). 
However, the ratio of the observed and true mean SFR is constant as a function of local galaxy density and is of the same order at any redshift. This implies that using our dataset we are likely over-estimating the mean observed SFR in the same way at any density without biasing the slope of the relation. Thus, our estimate of the SFR--density relation is rather robust despite the spectroscopic incompleteness. 
We use the dispersion estimated with this procedure to define the errors on the observed SFR--density relation at the corresponding density. This is possible because we find a correspondence between galaxy density field and galaxy parent halo mass both in the observed and simulated datasets, at least in the group regime.

\subsection{Is the SFR--density relation reversing at $z\sim 1$?}

The final point of our environmental approach focuses on understanding the disagreement between our findings and previous works claiming a reversal of the SFR--density relation at $z\sim 1$. The fairest comparison is with \cite{Elbaz2007} and \cite{popesso11}, since our dataset includes the sky regions covered by their dataset.

\begin{figure}
\centering
\includegraphics[width=\hsize,angle=180]{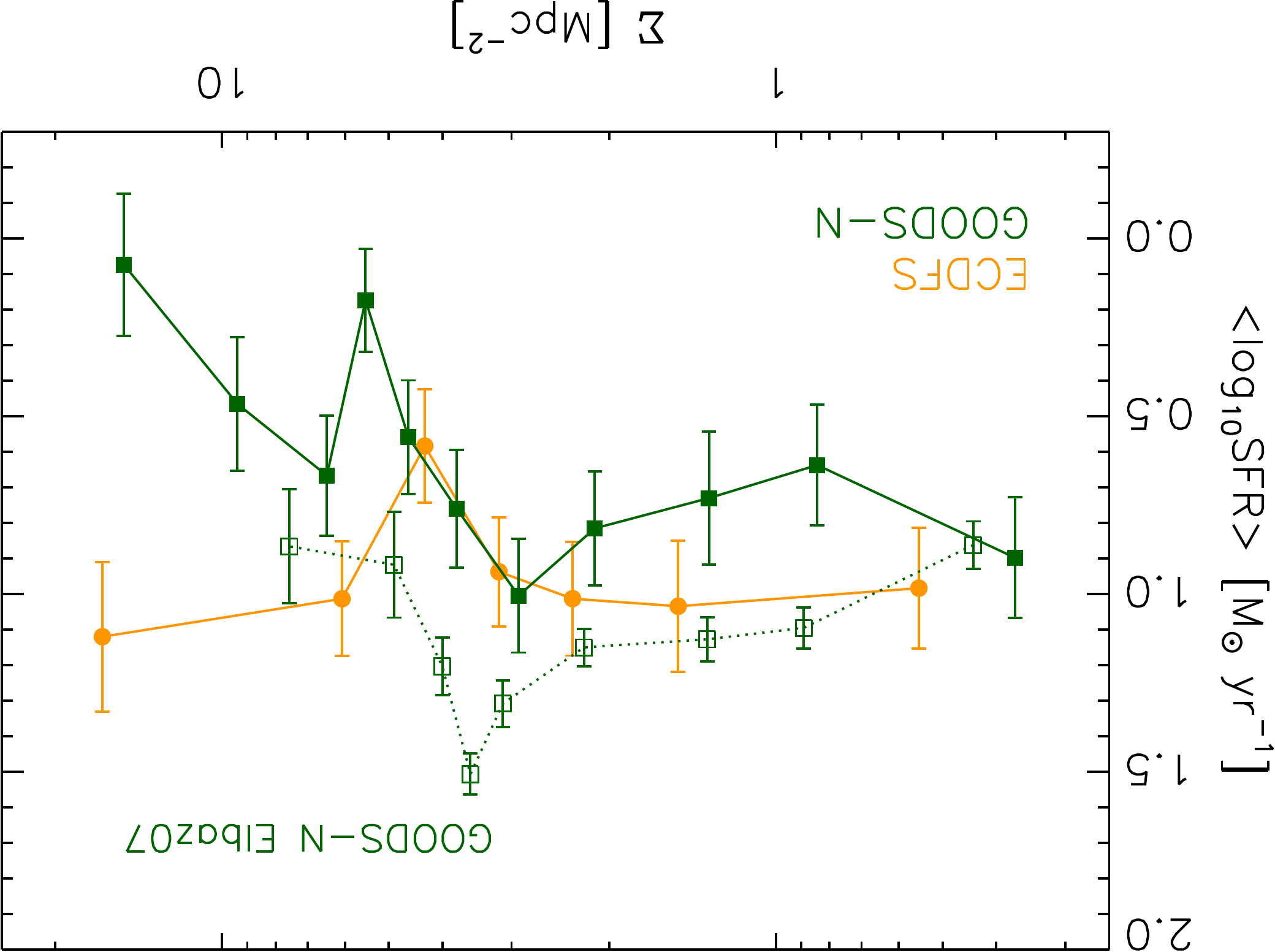}

\caption{SFR--density relation for all galaxies  with $\rm M_\star>10^{10.3}\,M_{\odot}$ at $z\sim1$. Galaxies in the ECDFS and GOODS-N fields are shown in orange and green, respectively. The open symbols connected by a dotted line show the SFR--density relation of \citet{Elbaz2007} for GOODS-North. 
}
\label{SFR_density_z1}
\end{figure}

Fig.~\ref{SFR_density_z1} shows the SFR--density relation at $0.8<z<1.2$ for the ECDFS and GOODS-N regions separately.  In the GOODS-N region we observe an anti-correlation between SFR and density with high significance, as confirmed by the Spearman test. We do not observe any relation between SFR and density in the ECDFS region which contains only a very poor group at $z=0.96$, differently from the GOODS-N region that comprises, in the same redshift bin, two very massive groups ($\rm M_{200}\sim 9\times 10^{13}\,M_\odot$). Errors are estimated as in Sec.~\ref{mock_errors}. 

Fig.~\ref{SFR_density_z1} also shows the relation of \cite{Elbaz2007} for the GOODS-N region with open symbols connected by a dotted line. 
We can compare our results with \cite{Elbaz2007} only qualitatively, since the definitions of the density parameter and of the galaxy sample differ considerably. In fact, \cite{Elbaz2007} include all galaxies with HST-ACS $z_{AB} < 23.5$ mag without any mass cut. Given the broad redshift range considered ($0.8<z<1.2$), this apparent magnitude cut corresponds to a difference of 0.75 mag from the lowest to the highest redshift limit, introducing a bias with respect to our physical stellar mass selection (see also \citealt{Cooper2010}). 
This could explain the offset between our SFR--density relation and that of \cite{Elbaz2007}. 
\cite{Elbaz2007}  find a positive correlation in GOODS-N up to the point in which the SFR reaches its maximum and then a rapid decline at higher density. They refer to this as a ``reversal'' of the SFR--density relation.  Even if we do not detect any reversal, we point out that the trend we observe for the GOODS-N field has a shape similar to that of \cite{Elbaz2011}. 

The analysis of the SFR--density relation of \cite{Elbaz2007} is based on the estimate of the mean SFR per density bin, rather than on statistical tests such as the Spearman test used in this work. In addition, the errors estimated in \cite{Elbaz2007} seem to be under-estimated with respect to ours. Indeed, Elbaz et al. (2007) use a bootstrap technique that can not take into consideration the effect of cosmic variance due to the relatively small fields considered in the analysis. 

\cite{popesso11} show that the use of PACS data provides a big advantage (with respect to the MIPS data) in measuring the unbiased SFR of AGN hosts, whose SFR could be enhanced with respect to non-active galaxies of similar $\rm M_\star$. Thus, given the high fraction of AGN (17\%) measured at least in the highly star-forming population of the GOODS-S and GOODS-N fields, \cite{popesso11} conclude that the reversal of the SFR observed by \cite{Elbaz2007} could be due to a bias introduced by the SFR of AGN host galaxies measured with MIPS data. 

In building the SFR--density relation, we are including all galaxies above $\rm 10^{10.3}~M_{\odot}$ with SFR much below the LIRG limit used by \cite{popesso11}. Taking advantage of the AGN sample of \cite{Shao2010} for the GOODS-N region and the AGN sample of \cite{Lutz2010} for the ECDFS, constructed with similar criteria and X-ray flux limits, we investigate whether AGN can bias our sample.  We observe an AGN fraction of 3-5\% in the ECDFS and GOODS-N region above our mass cut.
This fraction is much lower with respect to the work of \cite{popesso11}, who show that the fraction of AGN is much higher in highly IR luminous galaxies. Since we include galaxies spanning a wide range in SFR, the AGN fraction is diluted in our sample. If we remove the AGN from our 
sample, the significance of the SFR--density relation does not change at all, in agreement with \cite{Elbaz2011}. 

We conclude that the previously observed reversal of the SFR--density relation at $z\sim 1$ is most likely due to a combination of different effects: the galaxy sample selection, a rather high fraction of AGN in the selected sample and a possibly biased definition of the density parameter, which can hide a redshift dependence. 
In addition, we point out that the significance of this reversal is probably due to an under-estimate of the error on the mean SFR, since the cosmic variance is neglected.

\subsection{The ``dynamical'' approach}
\label{sfr_density_dynamical}

As shown in Fig.~\ref{density}, on the right of the dashed green line, where $\sim 90\%$ of group galaxies are located, there is still a large number of galaxies at densities comparable to groups but not associated with any extended emitting source identified by the X-ray catalogue of Finoguenov et al. (in prep.). Those galaxies are likely located in unbound large scale structures, such as filaments, or in dark matter haloes of lower mass with respect to the detection limits of the CDFS (Chandra Deep Field South) 4Ms \citep{Xue2011}. 

If the relative vicinity to other galaxies is the main driver in quenching the galaxy SF, we should not observe any difference in the level of SF activity between galaxies showing the same local galaxy density. If, instead, processes related to the dark matter halo play a stronger role, we should observe a difference in the level of SF activity between group galaxies and systems at high density but not related to massive dark matter haloes. 

To check this issue, we investigate the SFR--density relation with a novel ``dynamical'' approach. We distinguish between galaxies in three different environments: a) group members, as identified via dynamical analysis, b) ``filament-like'' galaxies identified as systems at the same density as group galaxies but not associated with any of the extended X-ray sources or to the \cite{Kurk2009} structure, c) isolated galaxies with local galaxy density $\Sigma < 4.5$ $\rm{Mpc}^{-1}$ (on the left-hand side of the green line of Fig.~\ref{density}), i.e. where we find a low fraction of group galaxies (8\%). We build a new version of the SFR--density relation by comparing the mean SFR in the three environments for all galaxies with $\rm M_\star>10^{10.3}~M_\odot$. This method allows us to isolate the contribution of 
 groups with halo mass $\rm 10^{13} \lesssim M_{200}/M_{\odot} \lesssim 2\times 10^{14}$ in the SFR--density relation.

\begin{figure*}
\centering
\includegraphics[width=\hsize]{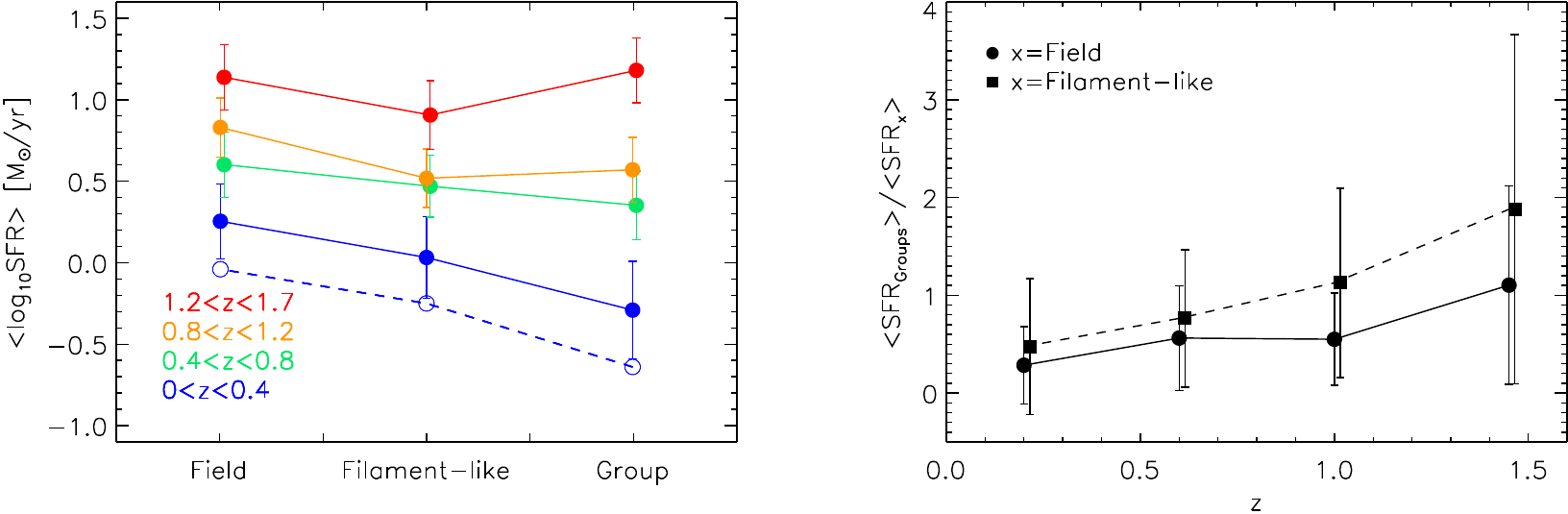}

\caption{Left: Comparison of mean SFR among group (within ${\rm 2 \times R_{200}}$), ``filament-like'' and field galaxies. Errors are derived using the mock catalogues of \citet{KW_millennium2007}, as explained in the text. The dashed line represents the mean SFR for the different environments at $0 < z < 0.4$ for all galaxies with $\rm M_\star> 10^9\,M_{\odot}$. Right: ratio between the $\rm \langle SFR \rangle$ of group galaxies with respect to field and ``filament-like'' galaxies as a function of redshift. In both panels data points are artificially shifted for clarity.}
\label{fig:plots_xbin1}
\end{figure*}

The left panel of Fig.~\ref{fig:plots_xbin1}, shows the SFR--density relation according to our new definition. We see a strong evolution with redshift of the mean SFR in groups (within ${\rm 2 \times R_{200}}$) with respect to the other two environments. Indeed, as shown in the right panel of Fig.~\ref{fig:plots_xbin1}, the ratio of $\rm \langle SFR\rangle $ is strongly evolving and it shows that the higher the redshift the lower the difference between the level of SF activity in groups and that in the field. This is consistent with the significance of the SFR--density (in its standard definition) anti-correlation decreasing with redshift (see Sec.~\ref{SFR_density_environmental}). 
The right panel of Fig.~\ref{fig:plots_xbin1} also shows the ratio between $\rm \langle SFR\rangle $ of groups and ``filament-like'' galaxies.  Although the errors are large, it is possible to appreciate how the ratio increases with redshift, with the SF activity of group galaxies being twice that of ``filament-like'' galaxies at $z\sim 1.6$. If this trend were real,  the structure at $z\sim1.6$ would provide some hints of the enhancement of the SF activity in groups with respect to filaments.  However, since the errors are quite large we can not draw any definitive conclusion.  

The evolution of the SF activity in different environments allows us to better understand the traditional SFR--density relation. 
In fact, the mix of galaxies in different environments, but at the same densities, hides the strong evolution observed in the left panel of Fig.~\ref{fig:plots_xbin1}. 
Our results also suggest that quenching processes related to a massive dark matter halo must play a decisive role in the strong evolution of the SF activity of group members with respect to galaxies in other environments.

\begin{figure}
\centering
\includegraphics[width=\hsize]{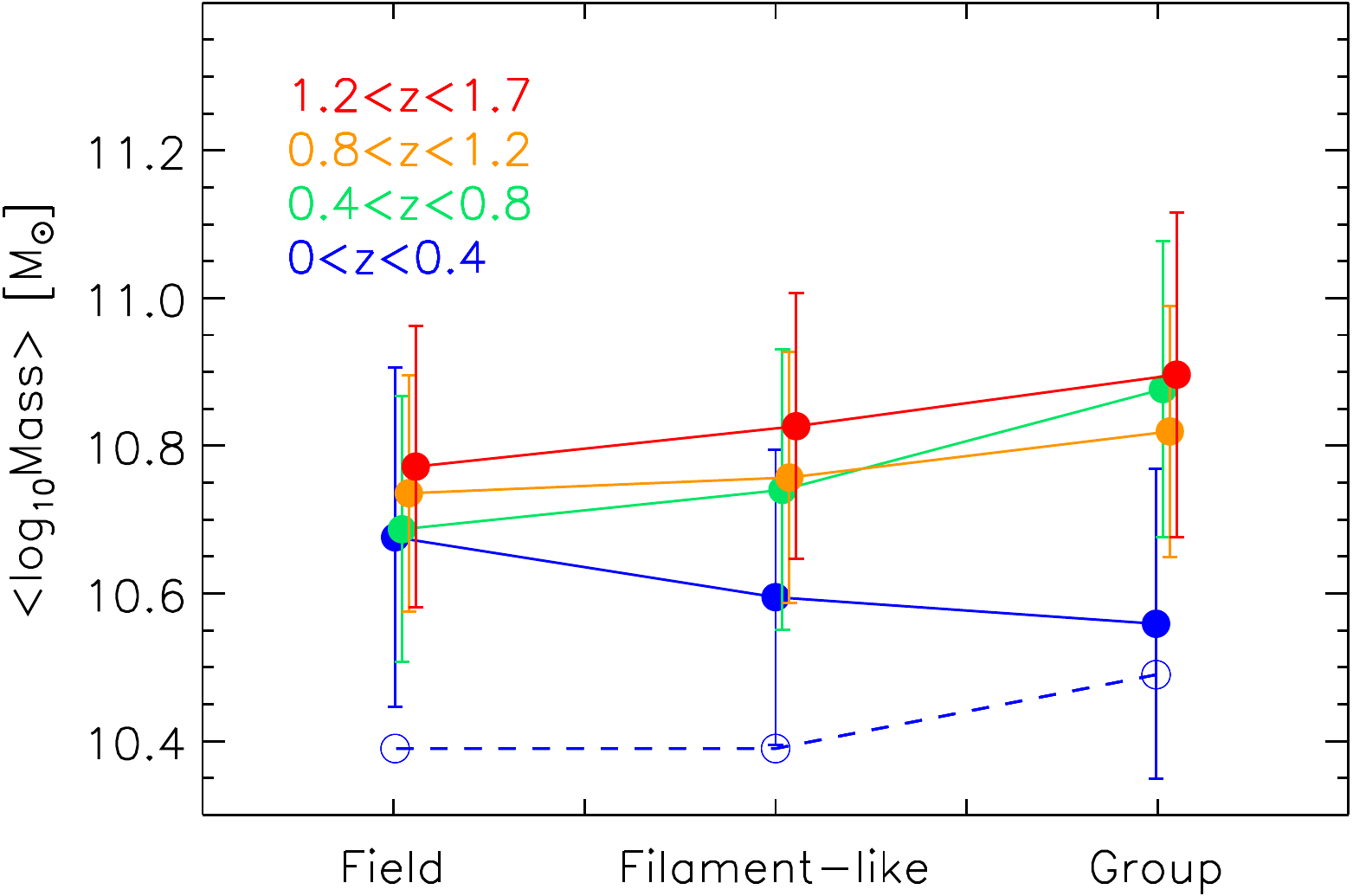} 
\caption{Comparison of mean stellar mass among group (within ${\rm 2 \times R_{200}}$), ``filament-like'' and field galaxies. Data points are artificially shifted for clarity. Errors are derived using the mock catalogues of \citet{KW_millennium2007}, as explained in the text. The dashed line represents the mean stellar mass for the different environments at $0 < z < 0.4$ for all galaxies with $\rm M_\star> 10^9\,M_{\odot}$. The normalisation of the dashed line is artificially increased to higher value to make it close to the blue solid line only for comparison.  }
\label{fig:plots_xbin2}
\end{figure}

We also check if the strong evolution of the newly defined SFR--density relation depends on a similar evolution of the $\rm M_\star$--density relation, using the same approach. Fig.~\ref{fig:plots_xbin2} shows the $\rm M_\star$--density relation in the usual four redshift bins according to our novel dynamical definition. 
The large errors do not allow us to see any  strong mass segregation. 
As in Fig.~\ref{mass_density}, the lowest redshift bin exhibits a different behaviour with respect to the
relation at higher redshifts. 
A mass cut at $\rm M_\star> 10^9\,M_{\odot}$ (dashed line and open symbols) allows us to highlight, once again, our spectroscopic bias on the lack of massive galaxies at low redshift.  With the same mass cut, the $\rm \langle SFR \rangle$ appears lower with respect to that derived with a higher mass cut at $\rm M_\star> 10^{10.3}\,M_{\odot}$ (dashed line and open symbols in Fig.~\ref{fig:plots_xbin1}), as expected by the MS evolution \cite[e.g.][]{Elbaz2007,Noeske2007a}. 
Thus, we conclude that, even with this approach, the strong difference between groups and low-density regime observed at $z < 0.8$ is likely not ascribable to a strong mass segregation.

For completeness, we also analysed the evolution of the sSFR--density relation. This relation evolves in the same way as the SFR--density relation, since the mass--density relation is only slightly evolving.

\subsubsection{Error analysis in the ``dynamical'' approach}
The errors on the mean SFR for group, ``filament-like'' and field galaxies are estimated in a similar way as in Sec.~\ref{mock_errors}. We randomly extract 100 catalogues (1000 regions with an area equal to the sum of the ECDFS and GOODS-N regions) in which we identify all haloes with masses between $\rm 10^{12.5}-10^{14} M_{\odot}$ and all their members. This information is obtained by linking the mock catalogues of \cite{KW_millennium2007} to the parent halo properties provided by the ``Friends-of-Friends'' algorithm \citep{Davis1985}   and the \cite{DeLucia2006} semi-analytic model tables of the Millennium database. In the same regions, we define the ``filament-like'' galaxies in the mock catalogues as the ones at the same density of the group galaxies but belonging to haloes with masses below $\rm 10^{12.5}~M_\sun$. Finally, field galaxies are defined as sources with densities below the threshold in the real dataset. 
We measure the mean SFR for group, ``filament-like'' and field galaxies ($\rm{SFR}_{incomplete, group}$, $\rm{SFR}_{incomplete, filament}$ and $\rm{SFR}_{incomplete, field}$, respectively) by using the galaxy members of each respective environment as in the observational dataset.

We measure in the same way the mean galaxy SFR ($\rm{SFR}_{real}$) for each population in the original (bias-free) \cite{KW_millennium2007} mock catalogues. We estimate, then, the difference  ${\rm \Delta SFR= log(SFR_{real})-log(SFR_{incomplete, i})}$ for each population, where $\rm{SFR}_{incomplete, i}$ is the mean SFR of the given population in the $i^{th}$ region.  The dispersion of the distribution of the residual ${\Delta{\rm{SFR}}}$ provides the error on our mean SFR. This error takes into account the bias due to incompleteness, the cosmic variance (due to the fact that we are considering small areas of the sky) and the uncertainty in the mean due to a limited number of galaxies per redshift bin. The bias introduced by the spectroscopic selection leads to an over-estimate of the mean SFR by the same amount, as expected, as in the case of the ``environmental approach''.  The same over-estimate is observed in each of the three populations. This is due to our assumption of a spatially-uniform sampling rate as the one guaranteed by the spectroscopic coverage of the ECDFS and GOODS-N fields.
The errors on the mean $\rm M_\star$ are estimated with the same procedure used for retrieving the errors on the mean SFR for each population.

\subsection{The $\rm SFR-M_\star$ plane in different environments}
\label{ms_analysis}

In this section we analyse the location of group, ``filament-like'' and low density (field) galaxies in the SFR--$\rm M_\star$ plane. This is done to identify the causes for the strong evolution of the SFR--density relation defined according to our ``dynamical'' definition.

\subsubsection{$\rm \Delta {MS}$ and $f_{QG}$ estimate}
\label{error_ms}
As already mentioned, \cite{Noeske2007a}, \cite{Elbaz2007}, \cite{Daddi2007a} and several other authors find a well defined sequence of star-forming galaxies in the SFR--$\rm M_\star$ plane from $z\sim 0$ to $z\sim2$. 
The relation shows a rather small scatter of 0.2-0.4 dex. 
The region below the main sequence is populated by quiescent galaxies (QG) in a scattered cloud, while only a small fraction (2\%) of outliers is found to be located above (by a factor of 4) the MS in the starburst region \citep[see][]{rodighiero10}. 

\cite{Noeske2007b} suggest that the same set of physical processes governs the SF activity in galaxies on this smooth sequence. If ``mass quenching'' \cite[e.g.][]{Peng2010} is the dominant mechanism for moving a galaxy off of the MS, the location of star-forming galaxies in high density regions should not be different from that of the bulk of the star-forming galaxies in other environments. Conversely, if the environment plays a role in the evolution of the galaxy's SF activity, the position of the group galaxies along or across the MS should be different with respect to the bulk of the star-forming galaxies. 

To shed light on this topic, we analyse the position of group, ``filament-like'' and field galaxies with respect to the MS in the ECDFS and GOODS regions and in four redshift bins.  In other words, we follow the behaviour of different environments defined in our dynamical approach (see Section \ref{sfr_density_dynamical}) in the SFR--$\rm M_\star$ plane.

Since the MS is well studied in the literature \citep[e.g.][]{Noeske2007a,Elbaz2007,Daddi2007a,Peng2010}, and the goal of this work is not to fit  this relation again, we use the best-fit relations available in the literature for the considered redshift bins. When no fit is available for a specific redshift bin, we interpolate the best-fit relations of the two closest redshift bins.

\begin{figure*}
\begin{center}
\includegraphics[width=0.49\hsize,]{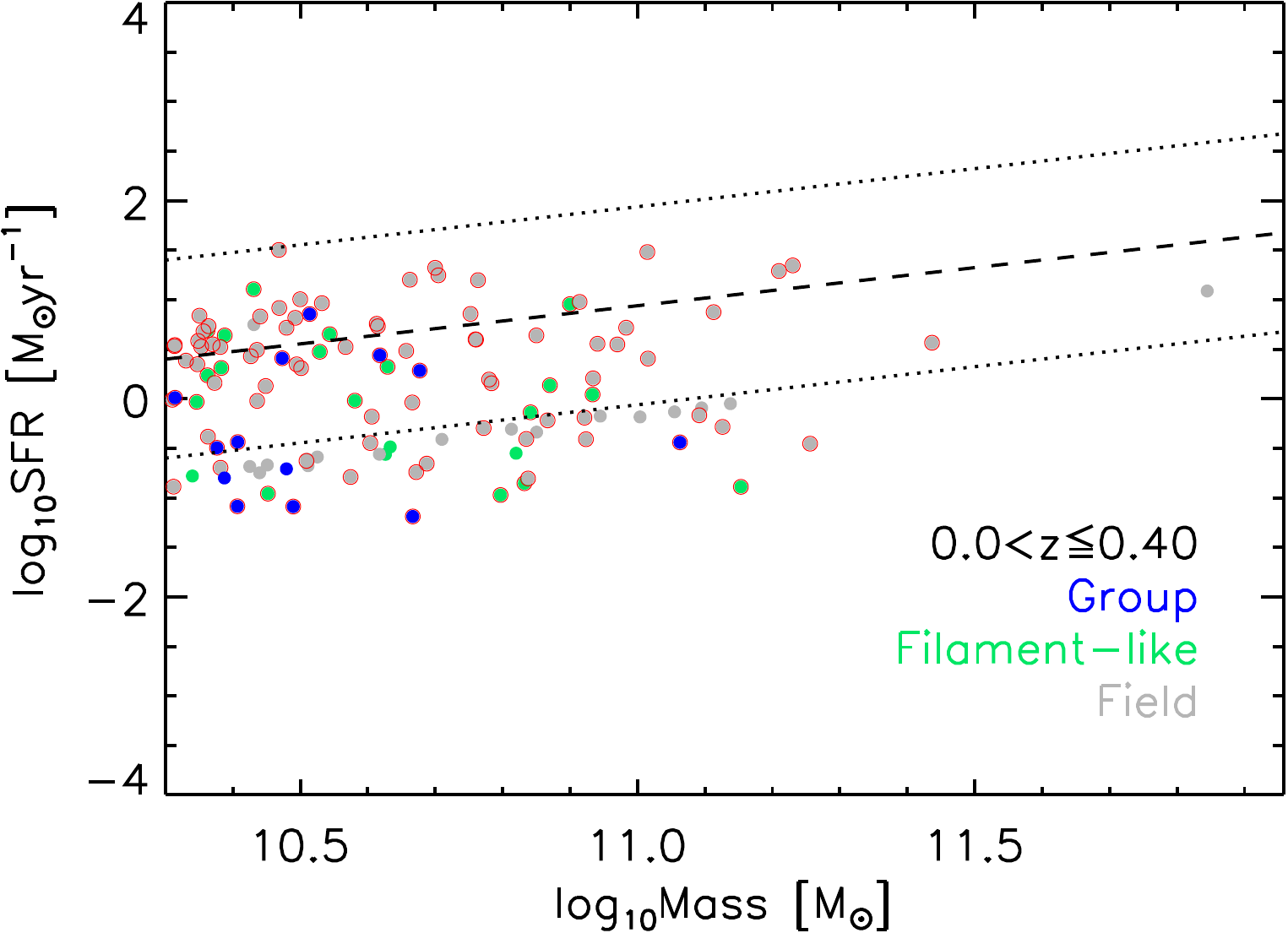}
\includegraphics[width=0.49\hsize,]{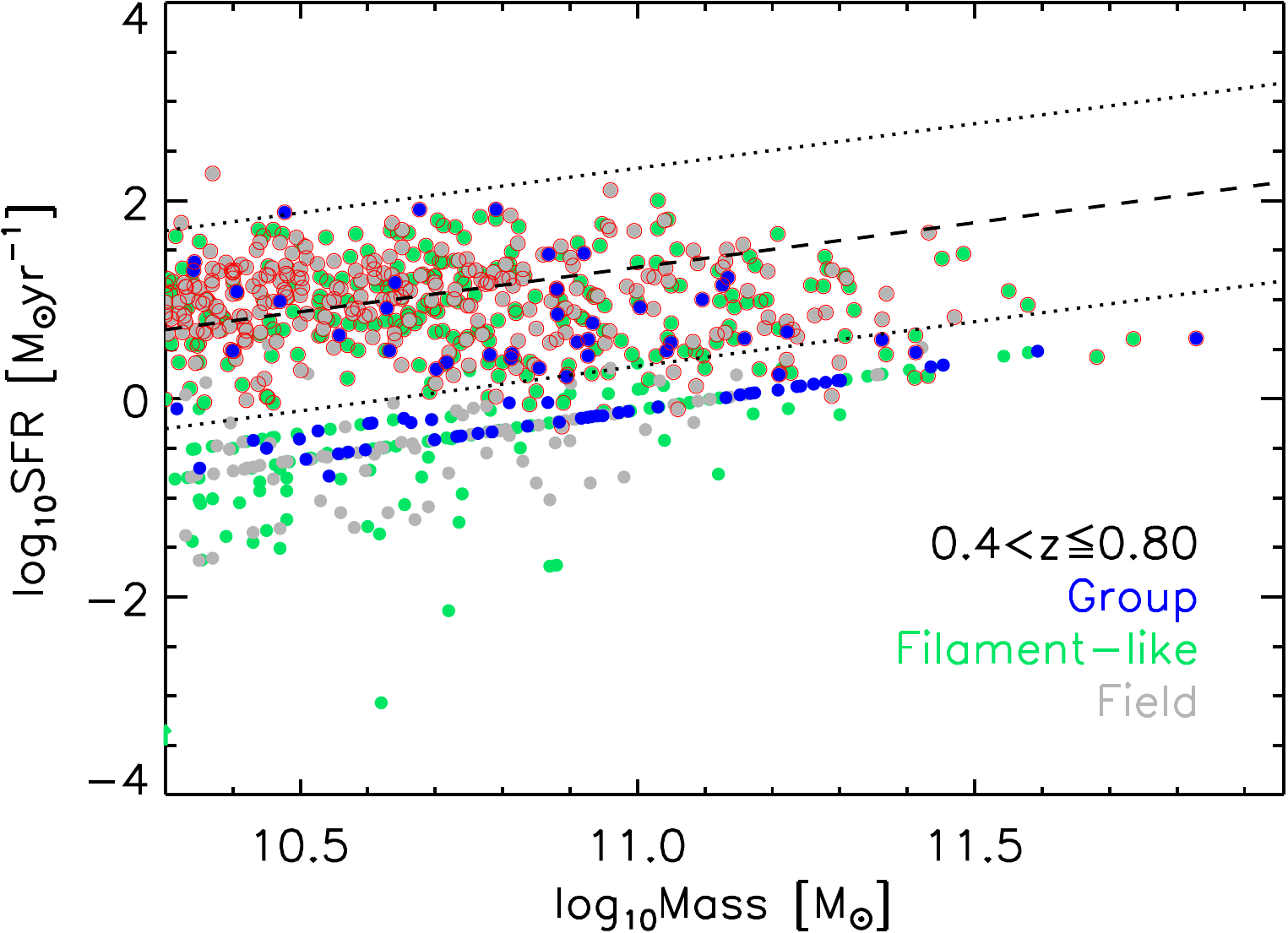}
\includegraphics[width=0.49\hsize,]{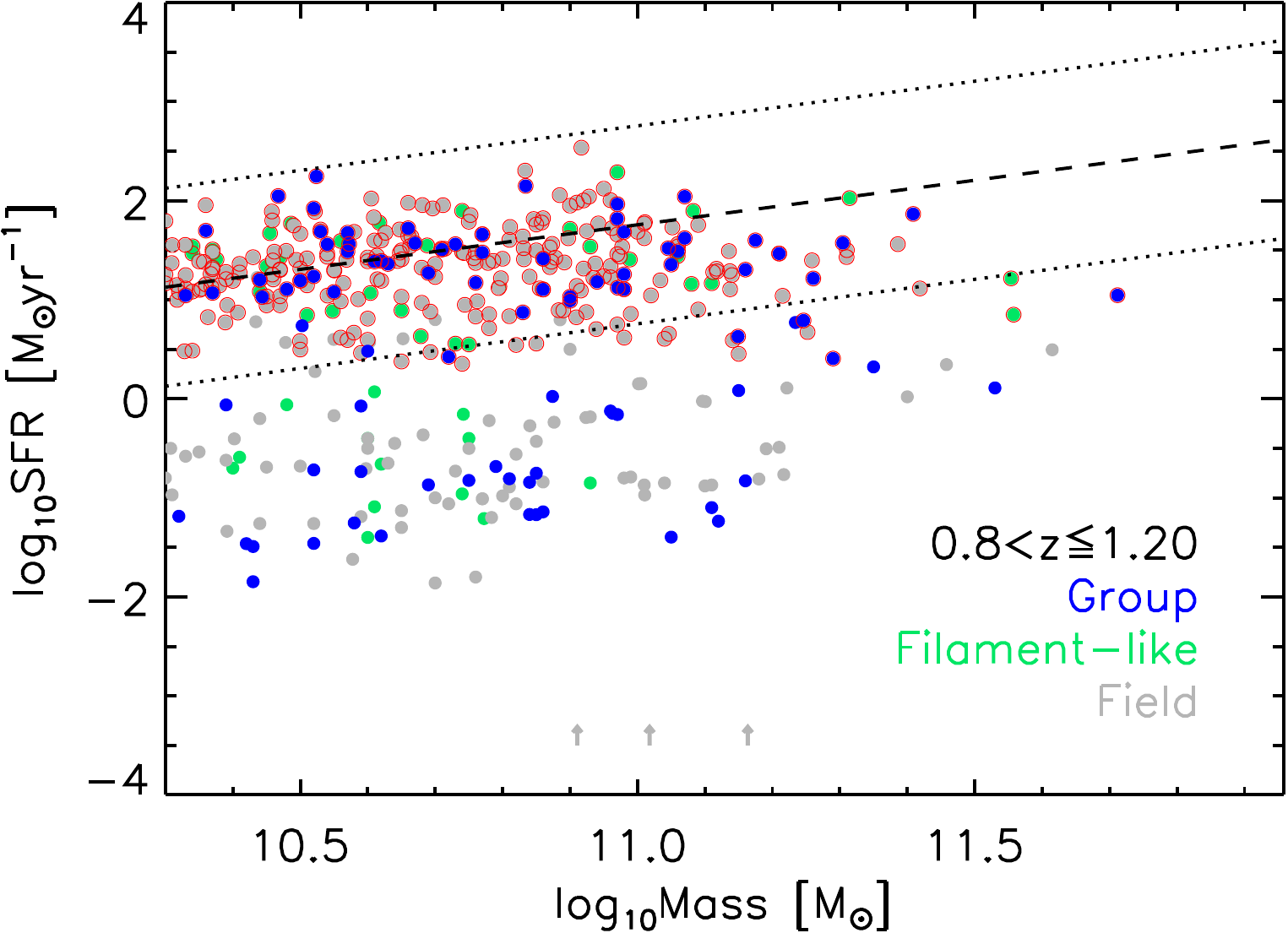}
\includegraphics[width=0.49\hsize,]{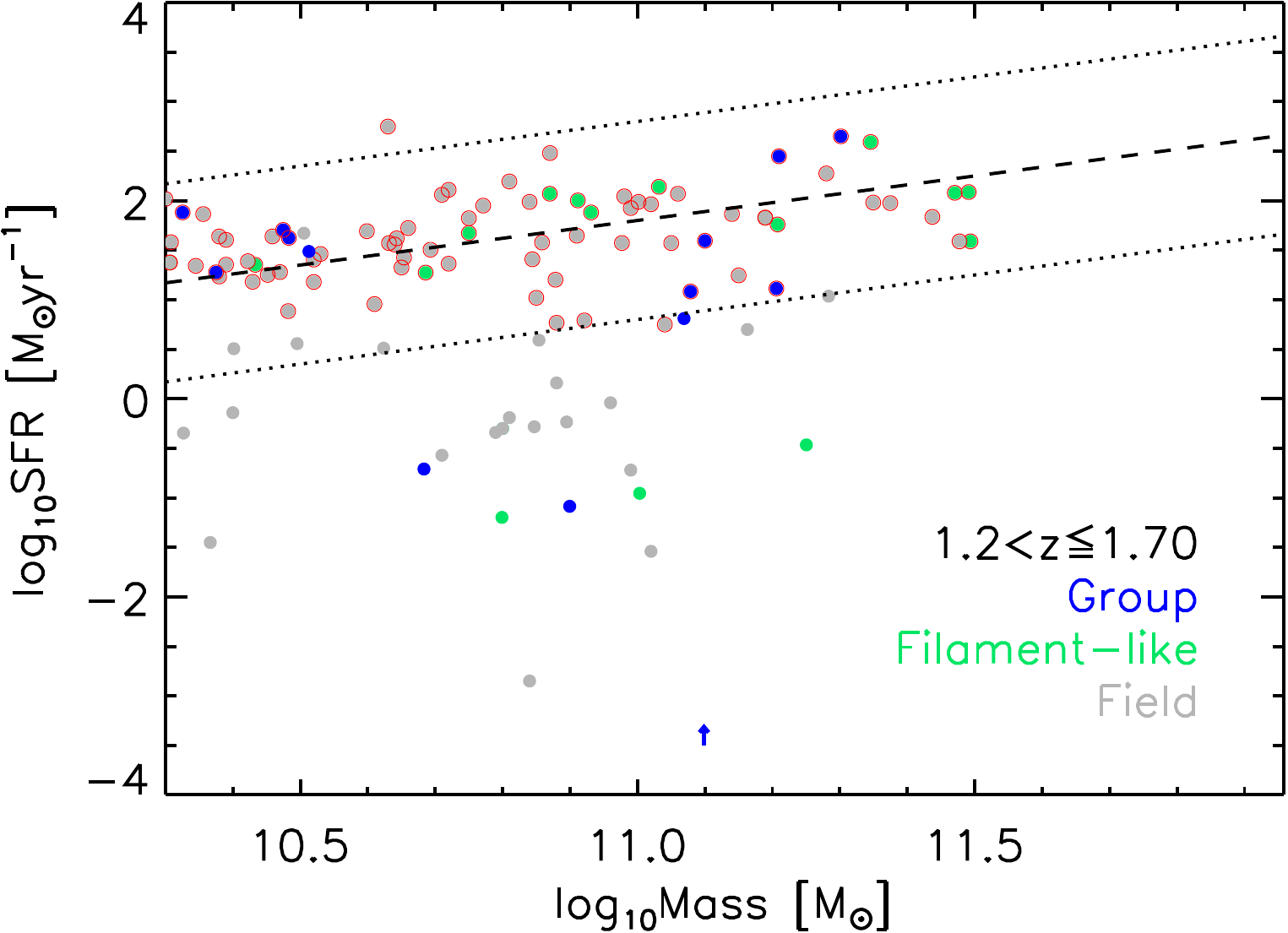}
\end{center}
\caption{$\rm SFR-M_\star$ diagrams for the different redshift bins considered in this work. We distinguish between group (blue filled circles) ``filament-like'' (green filled circles) and field (grey filled circles) galaxies. The empty red circles represent all galaxies detected in the IR bands. The dashed lines show the MS relations from the literature as explained in the text. MS galaxies are selected within the dotted lines, while QG are found below the dotted line, at lower SFR. The upward pointing arrows represent lower limits.}
\label{fig:sfr_m}
\end{figure*}

Fig.~\ref{fig:sfr_m} shows the SFR--$\rm M_\star$ planes for the different redshift bins and above the mass threshold considered in this work. In each plot, the grey filled circles show the field galaxy population, while green and blue filled circles represent the ``filament-like'' and group galaxy, respectively. We highlight with red empty circles all galaxies detected in the IR bands ($ >80\%$) by PACS and MIPS. 
On the other hand, the SFRs derived via the SED fitting technique are useful for defining the cloud of quiescent or low-star-forming galaxies below the MS. 
The issue with this SFR estimator is that it can create artificial discreet trails when plotting SFR and $\rm M_\star$. However, we point out that this does not affect our study. 
In the lowest redshift bin only a few galaxies populate the SFR--$\rm M_\star$ plane. This reflects the choice of giving a higher priority to spectroscopic targets like massive galaxies at high redshift (see Z13 and \citealt{Popesso2009}).

In order to define the MS in the four redshift bins, we use the equations already used in the literature which best represent our data.
In the $0<z\leq0.4$ redshift bin, we use a MS fit of \citet[their eq.~5]{Elbaz2007} based on SDSS star-forming galaxies at $z\sim 0$.
At $0.4<z\leq0.8$ we do not find a fit in the literature, thus, we interpolate the MS relation of \cite{Peng2010} at $z\sim0$ based on SDSS galaxies and that of \citet[their eq.~4]{Elbaz2007} at $z\sim1$ based on Spitzer MIPS detected galaxies.
The latter equation is used for the $0.8<z\leq1.2$ bin, with an offset\footnote{This offset does not affect at all our results, but it is necessary to better represent the MS field galaxy population.} of $log(SFR)=-0.16$.
As for the second redshift bin, the MS relation is not available in the literature for $1.2<z\leq1.7$ . Thus, we interpolate between the \citet[their eq.~4]{Elbaz2007} MS relation at $z\sim 1$  and the MS relation at $z\sim 2$ of \cite{Daddi2007a} based on UV data. 

In all cases we find a rather good agreement between our field galaxy distribution and the best-fit relations, with the mean of the distribution peaked at $\sim 0$ in the $\rm \Delta {MS}$ residual at all redshifts (left panel of Fig.~\ref{fig:ms_evolution}). 
We define $\rm{ \Delta MS=  log({SFR}_{observed}) - log({SFR}_{MS}})$ as the residual of the $\rm SFR-M_\star$ relation, where $\rm {SFR}_{observed}$ is the observed galaxy SFR and $\rm SFR_{MS}$ is the SFR predicted by the MS best-fit. We estimate $\rm {\Delta}MS$ for all the galaxies with mass above our mass cut ($\rm M_\star > 10^{10.3}~M_{\odot}$) and belonging to the three different environments in the usual redshift bins.  

At this point we identify and quantify the difference between the location across the MS of group galaxies in each bin with respect to the low density and ``filament-like'' galaxies. At all redshifts, the distribution of the $\rm \Delta {MS}$ residuals shows a  bimodal distribution: the Gaussian representing the MS location with a peak around 0, and a tail of quiescent/low-star-forming galaxies at low negative values of $\rm \Delta {MS}$. This distribution is reminiscent of the bimodal behaviour of the $u-r$ galaxy colour distribution observed by \cite{Strateva2001} in the SDSS galaxy sample. Following the example of \cite{Strateva2001}, we identify the minimum value of the valley between the MS Gaussian and the peak of the broader quiescent/low-star-forming galaxies distribution. At all redshifts, the value $\rm \Delta {MS}=-1$ turns out to be the best separation between the two galaxy populations. Since the observed scatter of the MS at any redshift varies between 0.2-0.4 dex \citep{Elbaz2007, Daddi2007a}, the limit at $\rm \Delta {MS}=-1$  should be consistent with a $3\sigma$ cut from the best-fit MS relation. We use this value to separate MS galaxies from quiescent/low-star-forming galaxies in the three considered environments. 

We measure the mean difference in SFR ($\rm \Delta {MS}$) from the MS location of the galaxy population in each environment, selecting only normally star forming galaxies in the range $\rm -1 \leq \Delta {MS} \leq 1$. By definition, $\rm \Delta {MS}$ should be consistent with 0 for the bulk of the MS galaxies. Thus, the mean of our Gaussian distribution centred around 0 confirms that our choice of the MS relation represents  well the mean of the normally star forming galaxies within our sample (but note the slope might not be so well represented, see the end of Section~\ref{sec:ms_qg_evolution}). 
We stress once again that given the depth of the PACS and {\it Spitzer}  MIPS observations of the ECDFS and GOODS fields, the MS is fully sampled (80\%) by IR-derived SFRs with very small (10\%) uncertainties. The SED fitting derived SFRs populate the region below the MS at $\rm \Delta {MS} <-1$, where we measure the quiescent galaxy fraction, $f_{QG}$. Thus, our estimate of the $\rm \Delta {MS}$ should not be affected by the large error (0.5-0.6 dex) in the determination of the SFR via SED fitting (see Z13 for details).

\subsubsection{Error estimates of $\Delta$MS and $f_{QG}$}

We estimate the error in $\rm \Delta {MS}$ ($f_{QG}$) with the same approach used in Sec.~\ref{mock_errors}. We use the usual 1000 regions with an area equal to the sum of the ECDFS and GOODS-N regions with a simulated spectroscopic incompleteness similar to our dataset. We identify group, ``filament-like'' and field galaxies in each region as explained in the error analysis of previous section. 

Our aim is to apply the same technique used to analyse the real dataset. Thus, we need the residual $\rm \Delta {MS}$ with respect to the MS relation to measure the mean distance from the MS at $\rm -1 < \Delta {MS} < 1$. 
However, the evolution of the MS predicted by the \cite{KW_millennium2007} catalogues is different from the one observed at the highest redshift considered in our work (see also \citealt{Elbaz2007}). Indeed, simulated star-forming galaxies, in particular at high redshift, tend to be less star-forming than in observations. Thus, the location of the MS using the mock catalogues at $z > 1$ tend to be below the observed MS in the same redshift bin. In order to cope with this problem we change the normalisation of the observed MS relation keeping the observed slope. Fitting the simulated MS provides similar results. 
In each area we measure the mean distance from the MS at $\rm -1 \leq \Delta {MS} \leq 1$ as is done in the real data ($\rm \langle \Delta {MS_{incomplete}}\rangle$).

We follow the same procedure in the original and complete \cite{KW_millennium2007}  mock catalogues by measuring  $\rm \Delta {MS_{real}}$. We measure, then,  the difference $\delta(\Delta {MS})=\rm \Delta {MS_{real}}-\Delta {MS_{incomplete, i}}$ for each population, where $\rm \langle \Delta {MS_{incomplete, i}}\rangle$ is the residual of the considered population in the $i^{th}$ region. 
The dispersion of the distribution of the residual $\rm \delta(\Delta {MS})$ provides the error on the observed $\rm \langle \Delta {MS}\rangle$. As in the previous case, this error takes into account the bias due to incompleteness, the cosmic variance and the uncertainty in the measure of the mean due to a limited number of galaxies per redshift bin.
We apply the same technique to estimate the $f_{QG}$ error in each of the three populations at different redshift.

\subsubsection{$\rm \Delta {MS}$ and $f_{QG}$ evolution} 
\label{sec:ms_qg_evolution}

\begin{figure*}
\begin{center}
\includegraphics[width=\hsize]{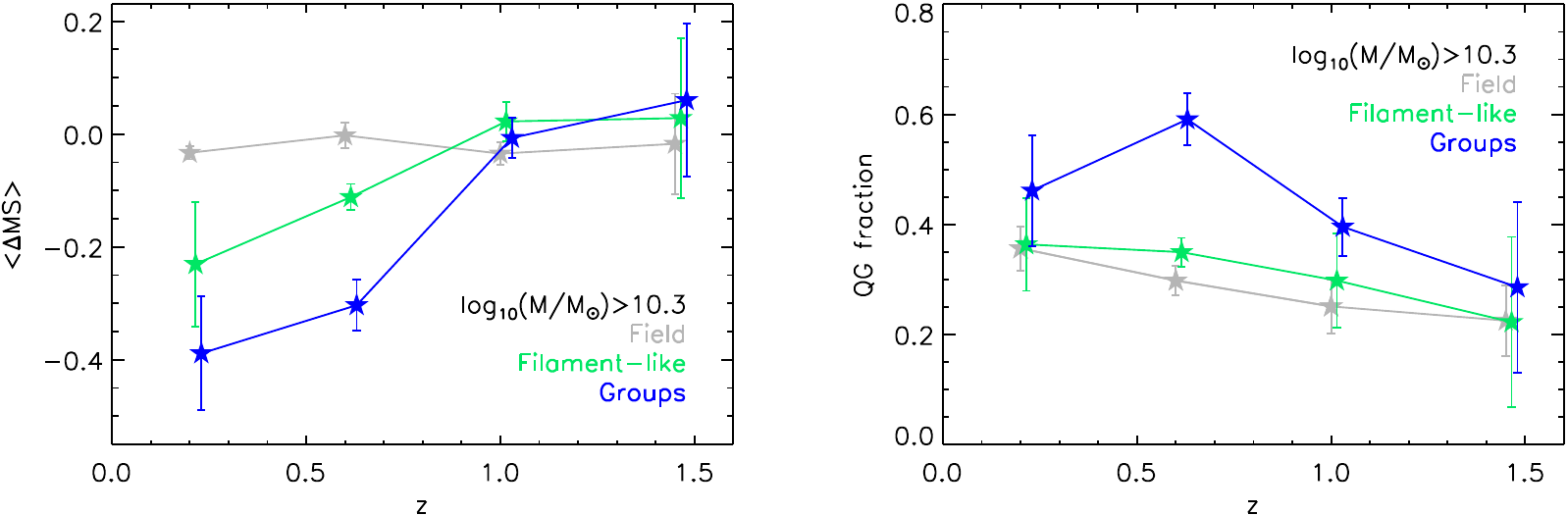}
\end{center}
\caption{Left: Evolution of the MS offset for group, ``filament-like'' and field star-forming galaxies with $\rm M_\star>10.3$. $\rm \Delta$MS represents the central value of the residuals with respect to the predicted MS for each redshift bin.  
Right: Evolution of the quiescent galaxy fraction ($f_{QG}$) for group (blue stars), ``filament-like'' (green stars) and field (grey stars) galaxies with  $\rm M_\star>10.3$.  We define ``quiescent'' all the galaxies with $\rm \Delta{\rm MS} <-1$. In both panels errors are derived using the mock catalogues of \citet{KW_millennium2007}, as explained in the text and data points are artificially shifted for clarity.
}
\label{fig:ms_evolution}
\end{figure*}

The left panel of Fig.~\ref{fig:ms_evolution} shows the evolution of the $\rm \langle \Delta {MS} \rangle$ for the MS galaxies in low density regions (grey stars and line), ``filament-like'' environments (green stars and line) and groups (blue stars and line) up to $z\sim 1.6$. In the first two redshift bins, the $\rm \langle \Delta MS \rangle$ of the star forming group galaxies is systematically below 0.  
At $z > 0.8$ the star-forming group galaxies are perfectly on sequence, consistently with the lower density environments. Moreover, the ``filament-like'' MS galaxies appear to be placed between the low density environment and the group galaxies. 

This result shows, for the first time, that at least below $z\sim 0.8$ the SF activity in star-forming group galaxies is lower than in the bulk of the star-forming galaxies. Here, we show that a certain amount of pre-processing (galaxies being pre-processed in groups before entering clusters, \citealt{Zabludoff_and_Mulchaey_1998}) happens even before star-forming galaxies enter the group environment. In fact, some quenching is already in place when galaxies fall along filaments or lower mass groups that could eventually merge to form more massive structures. Thus, the speed of the evolution of the SF activity in star-forming galaxies depends, at least since $z \sim 1$, on the galaxy environment. 

For completeness of the analysis, we also investigate the evolution of the galaxy-type mix for each environment. The galaxy-type mix is expressed  through the fraction of quiescent galaxies, $f_{QG}$. As already mentioned, we define as QG all those systems with $\rm \Delta {MS} < -1$, i.e. all the sources in the cloud below the MS. 
The right panel of Fig. ~\ref{fig:ms_evolution} shows the evolution of $f_{QG}$ in the three environments. Low density (grey stars and line) and ``filament-like'' galaxies (green stars and line) exhibit the same galaxy-type mix at any redshift and no evolution is observed in these environments at least in the mass 
range 
considered in our analysis. The galaxy-type mix in groups exhibits a higher $f_{QG}$ with respect to the other two environments, at any redshift. We note that the first redshift bin is affected by the spectroscopic selection function of our sample. The evolution of $f_{QG}$ is stronger in groups that in the other environments. In particular, we note that at $z \sim 0.8$ the fraction of quiescent galaxies is twice the mean fraction observed at high redshift.

The two panels of Fig.~\ref{fig:ms_evolution} show two different aspects of the role of environment in the evolution of galaxy SF activity. Some degree of partial quenching is observed as suggested by the ``environmental gradient'' as a function of distance from the MS. On the other hand, the right panel shows that the density is not responsible for the different galaxy-type mix. Indeed the group-  and ``filament-like'' regimes cover, by definition, the same range of local galaxy density. The main difference is that, in the group regime, galaxies likely belong to a massive ($\rm M_{200} \sim 2\times 10^{13} M_{\odot}$, see Z13 for the sample mass distribution) bound dark matter halo, while in the ``filament-like'' regime galaxies likely belong to unbound structures, such as filaments, or lower mass haloes. Thus, the different evolution of the galaxy-type mix of the two environments, similar in projected density but not in dynamical properties, indicates that the galaxy-type mix is connected indicates that a high $f_{QG}$ requires a massive parent dark matter halo rather than simply requiring an over density of galaxies.

\begin{figure*}
\begin{center}
\includegraphics[width=\hsize]{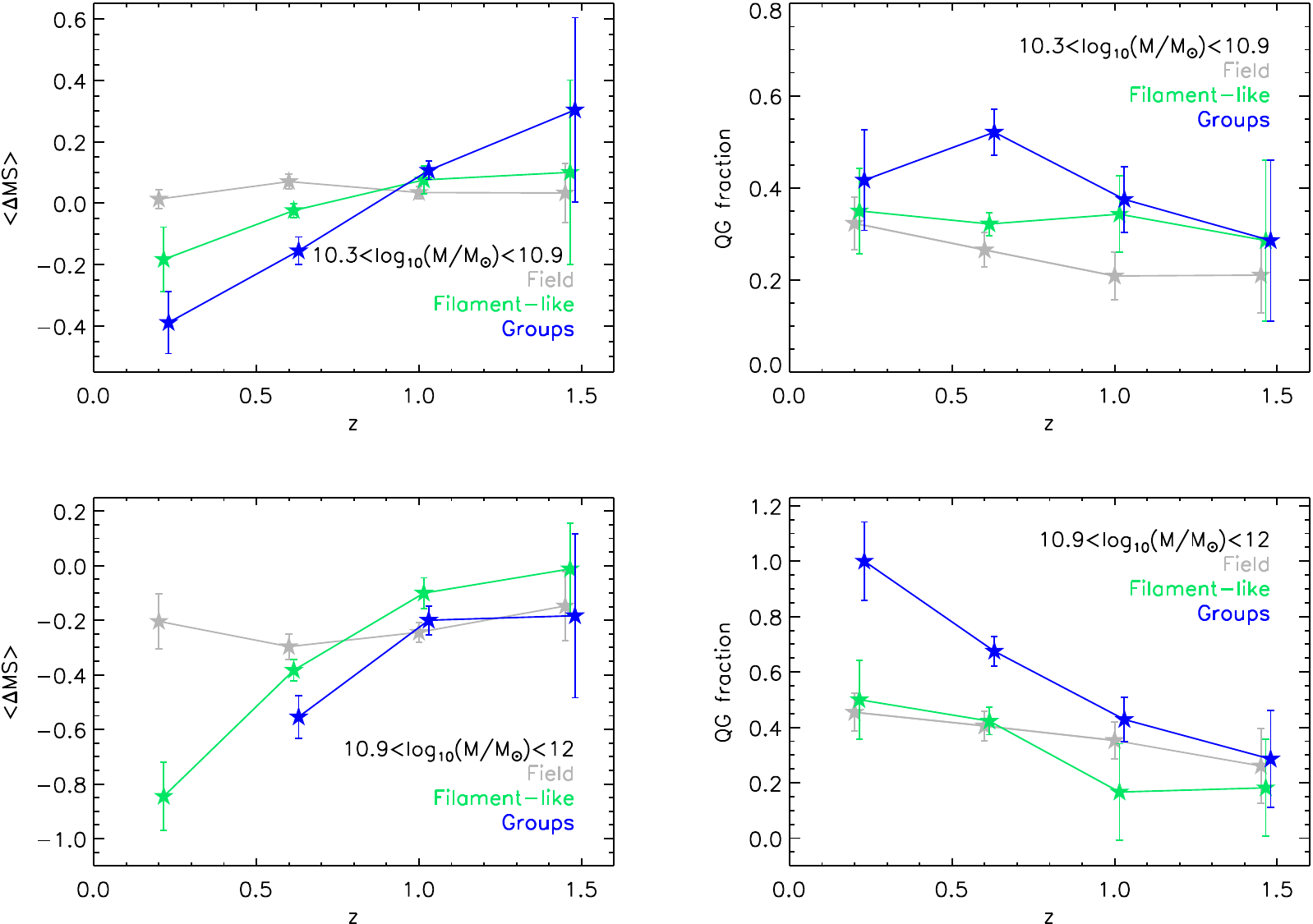}

\end{center}
\caption{Left column: Evolution of the MS offset for group, ``filament-like'' and field star-forming galaxies with $\rm 10.3<log(M_\star/M_\sun)<10.9$ and $\rm 10.9<log(M_\star/M_\sun)<12$ (top and bottom, respectively). 
Right column: Evolution of $f_{QG}$ for group (blue stars), ``filament-like'' (green stars) and field (grey stars) galaxies for the same range of stellar mass as in the left column. In all panels errors are derived using the mock catalogues of \citet{KW_millennium2007}, as explained in the text and data points are artificially shifted for clarity.  
}
\label{fig:ms_evolution_mbin}
\end{figure*}

In order to verify that our result is robust, we perform some sanity checks. This choice is driven by the distribution of galaxies in the $\rm SFR-M_\star$ plane. In fact, massive galaxies lie mainly below the MS (as defined, see Fig.~\ref{fig:sfr_m}), with the locus of galaxies apparently steepening towards lower SFR \cite[see e.g][]{Noeske2007b, Whitaker2012}. Thus, a non-zero $\rm \Delta MS$ could result from an environmental dependent distribution of mass coupled with a MS relation which does not represent the data at all masses.
The Kolmogorov-Smirnov (KS) test reveals that the mass distributions of star forming galaxies in different environments are consistent with one another in all redshift bins, except the second one. This could be a problem since a more significant population of the most massive galaxies is expected in groups compared to the field.
To check how our findings are affected by this issue, we investigate the evolution of $\rm \Delta MS$ and $f_{QG}$ in two different stellar mass bins: $\rm 10^{10.3}<M_\star/M_\sun<10^{10.9}$ and $\rm 10^{10.9}<M_\star/M_\sun<10^{12}$ (Fig.~\ref{fig:ms_evolution_mbin}). Reassuringly, we find results consistent with Fig.~\ref{fig:ms_evolution}. We note that the lowest redshift bin is populated by only a few galaxies and so we focus on higher redshifts. 

In general, in the low-mass bin we observe a similar trend in $\rm \Delta MS$ and $f_{QG}$ as in the total sample, while in the high-mass bin $\rm \Delta MS$ is systematically below 0 (bottom left panel of Fig.~\ref{fig:ms_evolution_mbin}). This illustrates that massive star-forming galaxies are not well represented by a linear MS relation (cf.~\citealt{Whitaker2012}). 
However, since the evolution of $\rm \Delta MS$ for the other environments is similar to the total sample, we conclude that the relative offset observed in Fig.~\ref{fig:ms_evolution} for ``filament-like'' and group galaxies is real. 
Also the evolution of $f_{QG}$ remains similar to the total sample in both stellar mass bins. Even though it is less populated than the low mass bin, the high-mass bin better highlights the different quenching in groups, ``filament-like'' environment and field.

%----------------------------------------------------------------------------------------

\section{Discussion}
\label{discussion}

\subsection{The SFR--density relation}

%----------------------------------------------------------------------------------------
We have investigated the SFR--density relation using two different approaches. One, more traditional, studies the relation between SFR and density for all galaxies, while the other (the ``dynamical'' approach) isolates the contribution of group galaxies with respect to other environments in the SFR--density relation.

Our results show that the SFR--density relation has progressively lower significance towards high redshift, but it does not reverse at $z\sim1$. In addition, a careful analysis of the biases due to the spectroscopic selection leads to the conclusion that we can also not exclude an anti-correlation at $z \gtrsim 1$.  The observed SFR--density anti-correlation at $z < 0.8$ is not simply ascribable to mass segregation (most massive galaxies are generally passive galaxies, low-mass galaxies are, on average, star-forming). Indeed, we observe only a mild mass segregation in any redshift bin.  

Our results seem to be at odds with \cite{Kauffmann_etal_2004} who find strong mass segregation at least in the local Universe. At higher redshift the effect has never been thoroughly analysed, except for the results of \cite{Scodeggio2009} and \cite{Bolzonella2010} who showed that already at $z \sim 1$ mass and galaxy density are coupled with the most massive galaxies segregated in the most dense environment.  We must note that our stellar mass 
cut ($\rm M_\star >10^{10.3}~M_\odot$) is rather high. Indeed, a lower mass cut ($\rm M_\star >10^{9}~M_\odot$) in the lowest redshift bin leads to a stronger mass segregation, although of small amplitude. This would be in agreement with the recent finding of \cite{Rasmussen2012}, who observe mass segregation within 10~\rvir~ for groups, by only considering low-mass galaxies.

Given the flattening of the SFR--density relation observed after excluding group galaxies from the sample, we conclude that group members are mostly responsible for the observed anti-correlation at $z < 0.8$. Thus, galaxies living in relatively massive dark matter haloes must have a suppressed mean SFR with respect to the field, at least up to $z \sim 0.8$. This is confirmed by the SFR--density relation analysed with our ``dynamical'' approach.

One of the most striking findings in our analysis is the lack of reversal of the SFR--density relation at $z\sim 1$. This result is at odds with recent findings. In particular, \cite{Elbaz2007} and \cite{Cooper2008} observe the reversal of the SFR--density relation at $z\sim1$ in the GOODS and the DEEP2  fields, respectively, using a spectroscopically defined density parameter. We have extensively compared our analysis with that of \cite{Elbaz2007} and \cite{popesso11}, since our dataset includes the sky regions analysed in their dataset (Fig.~\ref{SFR_density_z1}).  
In particular, we have considered the possibility that the fraction of AGN could affect the SFR estimate. In fact, since the fraction of AGN is found to be higher in groups as the redshift increases \citep{Tanaka2012, Georgakakis2007, Georgakakis2008} and the SFR of AGN host galaxies could be enhanced with respect to non active galaxies of similar stellar mass \citep{Santini2012, Rosario2013}, we use PACS data to measure, without biases, their SFR \citep{popesso11}. This can not be done using MIPS data or the [OII] doublet. 
Removing the AGN from our sample does not affect the significance of the SFR--density relation. This could be due to cosmic variance, since the fraction of AGN present in the GOODS-S field is higher (17\% in the highly star-forming galaxies, \citealt{popesso11}) with respect to ECDFS and GOODS-N (3-5\%).

Finally, we consider the possibility that the density definition itself could be responsible for the differences we observe.
In fact, as explained in Section~\ref{sec:galaxy_density}, our density estimate is based on a stellar mass cut. \cite{popesso2012} show that the density definition adopted by \cite{Elbaz2007}, based on a galaxy apparent magnitude cut ($z_{AB} < 23.5$ mag), could lead to a strong redshift bias.
Thus, we conclude that the previously 
observed reversal of the SFR--density relations is most likely due to the combination of different effects: the galaxy sample selection, high fraction of AGN and a possibly biased definition of the density parameter, which can hide a redshift dependence.

Our results are, instead, in agreement with \cite{Feruglio2010}, who find no dependence of the SFR and LIRG fraction on environment, arguing that the reversal, if any, must occur at $z>1$. According to \cite{Feruglio2010} the reversal found by \cite{Elbaz2007} and \cite{Cooper2008} might be due to the contribution of galaxies at lower stellar mass and SFR comprised in \cite{Elbaz2007} and \cite{Cooper2008} galaxy sample. However, since we consider a wide range of SFR and $\rm M_\star$, we disagree with this conclusion. The advantage of the \cite{Feruglio2010} study is the use of COSMOS data, which, due to its wide field, is less affected by cosmic variance (although \cite{Cooper2008} is the least impacted, covering a larger volume spread over 4 distinct fields). On the other hand, it must be noted that the authors use sources with both spectroscopic and photometric redshifts to define the density field. This could dilute any over-density present in the field. Our approach is, therefore, more rigorous in this sense. 

In our analysis we estimated the SFR using both IR data and multi-wavelength SED fitting. The latter is used for all galaxies undetected in the IR bands, i.e. below the MS. This SFR estimator gives a larger uncertainty with respect to the IR data, as shown in Z13. However, in case of an under-estimate of the $\rm SFR_{SED}$, i.e. for all galaxies that would be on the MS, we should have an IR detection, thus a more robust SFR estimate. 
On the other hand, in case $\rm SFR_{SED}$ over-estimates the real value, the signal could be washed out \citep{Cooper2010}. For example, any anti-correlation at $z>0.8$ could be hidden by the errors. In fact, as shown also by \cite{Wuyts2011}, the uncertainties on the SFR(SED) increase as a function of redshift.

We discuss our possible biases by using the mock catalogues of \cite{KW_millennium2007}. In all cases, we measure a $\rm \langle SFR \rangle$ higher than the one predicted by the simulations. This comparison also assures that we are not suffering from any bias in the slope of the SFR--density relation and, thus, that our results are robust.
Moreover, we note that in some cases the cosmic variance (or the big uncertainty in the $\rm SFR_{SED}$) could wash out the signal in the SFR--density anti-relation in the highest redshift bins.

The use of the standard approach for the study of the SFR--density relation, can be ineffective if the local galaxy density is not directly connected to the SF activity, due either to mass segregation or to SF quenching processes linked to galaxy-galaxy interactions. For this reason, we analyse the SFR--density relation with a novel ``dynamical'' approach. 
This technique allows us to separate the contribution to the highest galaxy density bins of groups and ``filament-like'' galaxies. 
This is not possible in a more classical ``environmental approach'' (although see \citealt{Wilman2010}). 

Our results show that the bulk of the SF is quenched in groups. This is what drives the trend of the SFR--density relation. The ``filament-like'' environment has a slower evolution in SF compared to the groups, thus the density (galaxy-galaxy interaction) itself can not be responsible for the bulk of the quenching.  
We note that we use a fairly large velocity window to compute densities, much larger than the typical velocity differences at which galaxy-galaxy interactions are effective. This means that there can be a certain dynamic range in the efficiency of our density estimates. In the X-ray groups, galaxies  have typically smaller velocity differences than chance projections, but larger differences than close pairs. Both the latter types contribute to ``filament-like'' environments.

This result is not necessarily inconsistent with Z13. In Z13, we show that the SF activity of galaxies is not affected by the local environment of groups, but here we find that it does depend on the global environment. In other words, the level of SF activity is generally low in groups (with respect to the other environments) even if it is independent from the group-centric distance (and from the density, as discussed in Z13). Rather than being inconsistent with Z13, this strengthens our results. In fact, we show, once again, that the density is not responsible for the bulk of the quenching (although we can not exclude that some quenching is happening for galaxy-galaxy interactions), but that  processes related to a massive dark matter halo are more effective.

The high SFR in the ``filament-like'' galaxies at low redshift, more consistent with the field than with the group $\rm \langle SFR \rangle$, is in agreement with the recent finding of \cite{Fadda+08} and \cite{Biviano2011}. They show that the filament around the super-cluster A1763 hosts the highest fraction of IR-emitting galaxies. Similarly, the ``filament-like'' region contains the highest total SFR per unit galaxy.  Our findings are also consistent with those of \cite{Porter_Raychaydhury2007}, who have used optical data to discover an enhanced star-forming activity among galaxies associated with filaments in the nearby Pisces-Cetus super-cluster.

\subsection{The $\rm SFR-M_\star$ plane in different environments}
In order to investigate the cause of the strong evolution of SF activity in our sample, we have studied the position of group, ``filament-like'' and field star-forming galaxies with respect to the MS galaxy population \citep{Elbaz2007, Noeske2007a, Daddi2007a, Peng2010}. 

Many works focus mainly on the study of the MS in field galaxies. For example,  according to  \cite{Noeske2007b}, this sequence suggests that the same small set of physical processes governs the SF activity in galaxies. 
Thus, if galaxy evolution is driven mainly by their $nature$, there should be no difference among MS star-forming galaxies, regardless of their environment. On the other hand, if a galaxy depends on the environment in which it lies, a group member should have a different level of SF with respect to the bulk of star-forming galaxies on the MS.

This last point reflects our main result from the analysis of the $\rm SFR-M_\star$ relation.  In particular, we have studied, for the first time, the location of galaxies in different environments on the $\rm SFR-M_\star$ plane. The evolution of $\rm \langle \Delta MS \rangle$ shows that, at least below  $z\sim 0.8$, the SF activity in group galaxies is quenched with respect to the bulk of star-forming galaxies. At earlier epochs, group, ``filament-like'', and field galaxies have comparable SF.  Interestingly, the density seems to play a role in the distance from the MS, since the filaments represent a somewhat intermediate environment in the evolution of $\rm \langle \Delta MS \rangle$. Therefore, we show, with high significance, that the speed of the evolution of SF activity in star-forming galaxies depends, at least since $z \sim 1$, on the galaxy environment, defined according to our ``dynamical'' approach.  
In addition, we find that the fraction of QG, $f_{QG}$, evolves faster in groups with respect to both filaments and field. The latter two environments show a similar evolution of $f_{QG}$. This confirms that quenching processes related to a rather massive dark matter halo ($\rm \gtrsim 2\times 10^{13} M_{\odot}$) are more efficient than those associated with a generally dense region. Thus, strangulation \citep{Larson1980}, ram pressure stripping   \citep{Gunn_Gott1972} and  harassment \citep{Moore1996}, are likely to be much more effective than the simple galaxy-galaxy interaction.

Our result is in contrast with the analysis of \cite{Peng2011}.  These authors  argue that central star-forming galaxies are equivalent to field galaxies. They claim that there is no difference in the main sequence relation of central and satellite galaxies. However, it is not clear how they discern between star-forming and passive galaxies. In fact, over their whole study, \cite{Peng2011} use the red/blue galaxy dichotomy to distinguish between passive/star-forming galaxies respectively. As shown by \cite{Woo2012}, about 30\% of the SDSS red sequence galaxies, identified in the colour-magnitude diagram, lie on the MS, which is also populated by green valley galaxies \citep{Rosario2013}. 
\cite{Whitaker2012} confirm this point, finding two different MS for blue and star-forming galaxies. This implies that selecting blue galaxies misses many red, dusty, star-forming sources. Moreover, \cite{Peng2011} use the catalogue of \citet[based on a Friends-of-Friends  algorithm]{Yang2007}  to explore the properties of group galaxies. 
As we already mentioned, optical selection of group is much more prone to projection effects than X-ray selection.  
In fact, the optically selected group catalogues do not contain virialized, relatively high-mass, X-ray emitting groups and include many more low-mass and unvirialized groups, as well as some pure projections. Finally, (part of) optically selected groups could be classified as ``filament-like'' galaxies, implying an  $f_{QG}$ evolution similar to that of field galaxies.

Recently, \cite{Rasmussen2012}, computing the SFR from UV emission for nearby group galaxies, found a MS broadly consistent but flatter than the MS of field galaxies at the same redshift. They argue that a flattening could be expected if the SFR of low-mass galaxies is suppressed in groups. At a median mass of $\rm log(M_\star/M_\odot)=9.63$, their MS predicts a mean sSFR which is $\sim 40\%$ lower than that expected for the field.  This could be consistent with our results, although we do not cover the same mass range.

Our findings support the pre-processing scenario (galaxies age in groups before entering clusters; \citealt{Zabludoff_and_Mulchaey_1998}).  This is consistent with the result of \cite{Wilman2008}, who find a strong suppression of SF activity in a sample of $z\sim0.4$ groups. They estimate a stronger quenching for galaxies with $\rm M_\star \gtrsim 10^{11}~M_\odot$, where the fraction of star-forming galaxies falls down to $\sim$12\%.  
Our results are also in agreement with an analogous analysis done by \cite{Bai2010} on the sample of 2dF groups. The authors show that the group star-forming galaxies are located below the field MS, but above the location of the bulk of cluster star-forming galaxies. 
This suggests that, although some pre-processing is present in groups, a stronger quenching must happen in more dense and massive systems, like clusters.
In this work we also show that a certain amount of pre-processing happens when galaxies are falling along the filaments, before they enter the group environment. However, halo-related processes seem to be more effective in quenching the star formation.

The pre-processing scenario is also supported by models.  \cite{DeLucia2012}, using semi-analytic models, show that the fraction of galaxies that can be pre-processed in a group-size halo of mass $\rm \sim 10^{13} ~ M_\odot$ is significant ($\sim 27\%$ which raise to $\sim44\%$ for galaxies with $\rm M\sim 10^{11}~M_\odot$). Furthermore, comparing observations with their theoretical predictions, they argue that satellite galaxies become passive after they have spent 5-7 Gyr in haloes more massive than $\rm M_{halo}\sim 10^{13} ~ M_\odot$.  Similarly, \cite{McGee2009}, using the stellar masses and merger trees produced by the semi-analytic galaxy catalogues, suggest that all clusters in their sample exhibit a significant fraction of their galaxies accreted through galaxy groups. For instance, they propose that this fraction is 40$\%$ for $\rm 10^{14.5}~M_{\odot}$ clusters at $z=0$ and only $\sim$ 25 $\%$ at higher redshifts ($z\sim1.5$). Our results show qualitative agreement with this prediction. 
Conversely, \cite{Berrier_etal_2009}, using cosmological $\Lambda$CDM N-body simulations, suggest that on average, $\sim$70\% of cluster galaxies fall into the cluster potential directly from the field. On the other hand, less than $\sim$12\% of cluster galaxies are accreted as members of groups with five or more galaxies.

The pre-processing scenario is also reflected in our analysis on the QG fraction in groups, ``filament-like'' environments and field. Our findings suggest that these environments have a different galaxy-type mix up to $z\sim 1$, with groups being the most efficient at quenching the SF. At higher redshift, the galaxy population of groups, filaments and field is similar.

Our results find support in several works in the literature. For example, \cite{Kovac2010} show that galaxy star-formation and colour transformation rates are higher in the group regions than in lower density areas at $z \sim 1$. In addition, \cite{Presotto2012} suggest that galaxy colours are particularly affected by the group environment (with respect to the field) on short time-scales in a redshift range $0.2<z<0.8$. Finally, \cite{Iovino2010} and \cite{Gerke2007} show that the group galaxy population becomes bluer as the redshift increases, but it maintains a systematic difference with respect to the global galaxy population, and an even larger difference with respect to the isolated galaxy population.

\section{Summary and conclusions}
\label{summary_conclusions}
In this work we have investigated the SFR--density relation in different environments up to $z\sim1.6$. We have used multi-wavelength data from the ECDFS and GOODS fields to study the evolution of SF activity in four redshift bins. Moreover, the use of deep MIPS 24~\um and {\it Herschel}  PACS data has assured an accurate estimate of the SFR for all detected IR sources (in particular MS galaxies). This rich dataset has enabled the use of two different approaches to investigate the evolution of the SFR--density relation: an ``environmental'' approach, which is the traditional method used in the literature, and a novel ``dynamical'' approach, which splits the sample into group, ``filament-like'' and field galaxies.

By studying the SFR--density relation in the standard way, we have found an anti-correlation up to $z\sim 0.8$ but no correlation at higher redshift. Although the significance found by the Spearman test decreases as the redshift increases, we did not observe any reversal of the SFR--density relation. 
After checking for the presence of biases using the mock catalogues of \cite{KW_millennium2007}, we have verified that we have constantly over-estimated the values of SFR at all redshifts, due to the spectroscopic incompleteness of our catalogues, but that the slope of our SFR--density relation is not affected by any bias. We have also found that the role of AGN is rather marginal in shaping the relation and that the anti-correlation at $z < 0.8$ is dominated by spectroscopic group members.  Since our galaxy sample shows only a mild mass segregation at any redshift bin, we conclude that the SFR--density relation is not driven by a strong mass segregation.

By using the ``dynamical'' definition of environment, we have asserted that the bulk of quenching happens in groups. Indeed, group spectroscopic members show a much lower mean SFR than galaxies at similar density but not belonging to bound structures, at least up to $z \sim 1$. On the other hand, galaxies in unbound structures exhibit a similar evolution of SFR as field isolated galaxies. Group galaxies only reach the same level of SF activity as field galaxies at $z > 1$. However, even with this alternative approach, we have not detected any significant SFR--density reversal. Thus, we conclude that group galaxies experience a much faster evolution with respect to galaxies in other environments. In addition, the strong difference in the evolution 
of the group galaxies with respect to non-group galaxies at similar density (i.e. ``filament-like'') reveals that processes related to the presence of a massive dark matter halo (ram pressure stripping, strangulation, harassment) must be dominant in the suppression of the SF activity in group galaxies below $z\sim1$. On the other hand, purely density-related processes (close encounters, tidal tripping) play a secondary role in the quenching.

In order to understand the cause of the faster evolution in group galaxies, we have also studied the location of group, ``filament-like'' and field galaxies in the $\rm SFR-M_\star$ plane. This has been done to identify if the lower $\rm \langle SFR  \rangle$ in groups at $z < 1$ with respect to field galaxies is due to a general quenching of the SF in all galaxies or to a faster evolution of the galaxy-type mix. We have found that the MS of group galaxies is offset with respect to that of field galaxies up to $z\sim 0.8$, i.e. it is shifted towards lower SFRs. At higher redshift the star-forming group galaxies are on sequence. ``Filament-like'' galaxies occupy a halfway position between groups and field. This suggests that both the density- and halo-related processes are playing a role in quenching the SF activity of actively star-forming galaxies, but that density seems to play a secondary role. Interestingly, the QG fraction evolves faster in groups than in the other two environments up to $z\sim 0.8$, beyond which the fractions are comparable. 
We conclude that the strong evolution observed in the SFR--density relation, analysed in the dynamical approach, is likely to be the driver of the different galaxy type mix in groups across cosmic epochs.

\section*{Acknowledgements}
We thank the anonymous referee for her/his constructive comments.

FZ acknowledges the support from and participation in the International Max-Planck Research School on Astrophysics at the Ludwig-Maximilians University.

We would like to thank Rob Yates for reading the paper and providing useful comments.

MT gratefully acknowledges support by KAKENHI No. 23740144.

FEB acknowledges support from Basal-CATA (PFB-06/2007), CONICYT-Chile (under grants FONDECYT 1101024, ALMA-CONICYT 31100004, and Anillo ACT1101), and {\it Chandra} X-ray Center grant SAO SP1-12007B.

PACS has been developed by a consortium of institutes led by MPE 
(Germany) and including UVIE (Austria); KUL, CSL, IMEC (Belgium); CEA, 
OAMP (France); MPIA (Germany); IFSI, OAP/AOT, OAA/CAISMI, LENS, SISSA 
(Italy); IAC (Spain). This development has been supported by the funding 
agencies BMVIT (Austria), ESA-PRODEX (Belgium), CEA/CNES (France),
DLR (Germany), ASI (Italy), and CICYT/MCYT (Spain).

This research has made use of NASA's Astrophysics Data System, of NED,
which is operated by JPL/Caltech, under contract with NASA, and of
SDSS, which has been funded by the Sloan Foundation, NSF, the US
Department of Energy, NASA, the Japanese Monbukagakusho, the Max
Planck Society, and the Higher Education Funding Council of England.
The SDSS is managed by the participating institutions
(www.sdss.org/collaboration/credits.html).

This work has been partially supported by a SAO grant SP1-12006B grant to UMBC.

\bsp

\label{lastpage}


\begin{thebibliography}{99}
\bibitem[\protect\citeauthoryear{Arnouts et al.}{2001}]{arnoutsetal2001} Arnouts S., et al., 2001, A\&A, 379, 740 
\bibitem[\protect\citeauthoryear{Bai et al.}{2010}]{Bai2010} Bai L., Rasmussen J., Mulchaey J.~S., Dariush A., Raychaudhury S., Ponman T.~J., 2010, ApJ, 713, 637 

\bibitem[\protect\citeauthoryear{Balestra et al.}{2010}]{balestraetal2010} Balestra I., et al., 2010, A\&A, 512, A12 

\bibitem[\protect\citeauthoryear{Barger, Cowie, \& Wang}{2008}]{Barger2008} Barger A.~J., Cowie L.~L., Wang W.-H., 2008, ApJ, 689, 687 

\bibitem[\protect\citeauthoryear{Berrier et al.}{2009}]{Berrier_etal_2009} Berrier J.~C., Stewart K.~R., Bullock J.~S., Purcell C.~W., Barton E.~J., Wechsler R.~H., 2009, ApJ, 690, 1292 

\bibitem[\protect\citeauthoryear{Biviano et al.}{2011}]{Biviano2011} Biviano A., Fadda D., Durret F., Edwards L.~O.~V., Marleau F., 2011, A\&A, 532, A77 


\bibitem[\protect\citeauthoryear{Bolzonella et al.}{2010}]{Bolzonella2010} Bolzonella M., et al., 2010, A\&A, 524, A76 


\bibitem[\protect\citeauthoryear{Bruzual \& Charlot}{2003}]{bruzual_charlot2003} Bruzual G., Charlot S., 2003, MNRAS, 344, 1000 

\bibitem[\protect\citeauthoryear{Cardamone et al.}{2010}]{Cardamoneetal2010} Cardamone C.~N., et al., 2010, ApJS, 189, 270 
\bibitem[\protect\citeauthoryear{Chabrier}{2003}]{Chabrier2003} Chabrier G., 2003, PASP, 115, 763 

\bibitem[\protect\citeauthoryear{Cooper et al.}{2005}]{Cooper2005} Cooper M.~C., Newman J.~A., Madgwick D.~S., Gerke B.~F., Yan R., Davis M., 2005, ApJ, 634, 833 
\bibitem[\protect\citeauthoryear{Cooper et al.}{2008}]{Cooper2008} Cooper M.~C., et al., 2008, MNRAS, 383, 1058 
\bibitem[\protect\citeauthoryear{Cooper et al.}{2010}]{Cooper2010} Cooper M.~C., et al., 2010, MNRAS, 409, 337 
\bibitem[\protect\citeauthoryear{Cooper et al.}{2012}]{Cooper2011} Cooper M.~C., et al., 2012, MNRAS, 425, 2116 
\bibitem[\protect\citeauthoryear{Daddi et al.}{2007}]{Daddi2007a} Daddi E., et al., 2007, ApJ, 670, 156 

\bibitem[\protect\citeauthoryear{Davis et al.}{1985}]{Davis1985} Davis M., Efstathiou G., Frenk C.~S., White S.~D.~M., 1985, ApJ, 292, 371 


\bibitem[\protect\citeauthoryear{De Lucia et al.}{2006}]{DeLucia2006} De Lucia G., Springel V., White S.~D.~M., Croton D., Kauffmann G., 2006, MNRAS, 366, 499 

\bibitem[\protect\citeauthoryear{De Lucia et al.}{2012}]{DeLucia2012} De Lucia G., Weinmann S., Poggianti B.~M., Arag{\'o}n-Salamanca A., Zaritsky D., 2012, MNRAS, 3042 

\bibitem[\protect\citeauthoryear{{Dressler}}{{Dressler}}{1980}]{Dressler_1980} Dressler A.,  1980, ApJ, 236, 351


\bibitem[\protect\citeauthoryear{Elbaz et al.}{2007}]{Elbaz2007} Elbaz D., et al., 2007, A\&A, 468, 33 

\bibitem[\protect\citeauthoryear{Elbaz et al.}{2011}]{Elbaz2011} Elbaz D., et al., 2011, A\&A, 533, A119 



\bibitem[\protect\citeauthoryear{Fadda et al.}{2008}]{Fadda+08} Fadda D., Biviano A., Marleau F.~R., Storrie-Lombardi L.~J., Durret F., 
2008, ApJ, 672, L9 

\bibitem[\protect\citeauthoryear{Fazio et al.}{2004}]{Fazio2004} Fazio G.~G., et al., 2004, ApJS, 154, 10 

\bibitem[\protect\citeauthoryear{Feruglio et al.}{2010}]{Feruglio2010} Feruglio C., et al., 2010, ApJ, 721, 607 

\bibitem[\protect\citeauthoryear{Finoguenov et al.}{2007}]{Finoguenov2007} Finoguenov A., et al., 2007, ApJS, 172, 182 


\bibitem[\protect\citeauthoryear{Finoguenov et al.}{2009}]{Finoguenov2009} Finoguenov A., et al., 2009, ApJ, 704, 564 

\bibitem[\protect\citeauthoryear{Gerke et al.}{2007}]{Gerke2007} Gerke B.~F., et al., 2007, MNRAS, 376, 1425 

\bibitem[\protect\citeauthoryear{Georgakakis et al.}{2007}]{Georgakakis2007} Georgakakis A., et al., 2007, ApJ, 660, L15 

\bibitem[\protect\citeauthoryear{Georgakakis}{2008}]{Georgakakis2008} Georgakakis A., 2008, AN, 329, 174 


\bibitem[\protect\citeauthoryear{George et al.}{2011}]{George2011} George M.~R., et al., 2011, ApJ, 742, 125 
\bibitem[\protect\citeauthoryear{George et al.}{2012}]{George2012} George M.~R., et al., 2012, ApJ, 757, 2 

\bibitem[\protect\citeauthoryear{Giavalisco et al.}{2004}]{Giavalisco2004} Giavalisco M., et al., 2004, ApJ, 600, L93 

\bibitem[\protect\citeauthoryear{G{\'o}mez et al.}{2003}]{Gomez2003} G{\'o}mez P.~L., et al., 2003, ApJ, 584, 210 
\bibitem[\protect\citeauthoryear{Grazian et al.}{2006}]{grazian06} Grazian A., et al., 2006, A\&A, 449, 951, 1 

\bibitem[\protect\citeauthoryear{Gunn \& Gott}{1972}]{Gunn_Gott1972} Gunn J.~E., Gott J.~R., III, 1972, ApJ, 176, 1 

\bibitem[\protect\citeauthoryear{Ideue et al.}{2009}]{Ideue2009} Ideue Y., et al., 2009, ApJ, 700, 971 
\bibitem[\protect\citeauthoryear{Ideue et al.}{2012}]{Ideue2012} Ideue Y., et al., 2012, ApJ, 747, 42 

\bibitem[\protect\citeauthoryear{Ilbert et al.}{2006}]{ilbertetal2006} Ilbert O., et al., 2006, A\&A, 457, 841 
\bibitem[\protect\citeauthoryear{Ilbert et al.}{2010}]{ilbertetal2010} Ilbert O., et al., 2010, ApJ, 709, 644 
\bibitem[\protect\citeauthoryear{Iovino et al.}{2010}]{Iovino2010} Iovino A., et al., 2010, A\&A, 509, A40 

\bibitem[\protect\citeauthoryear{Kauffmann et~al.}{2004}]{Kauffmann_etal_2004} Kauffmann G., White S.~D.~M., Heckman T.~M., M{\'e}nard B., Brinchmann J., Charlot S., Tremonti C., Brinkmann J., 2004,   MNRAS, 353, 713

\bibitem[\protect\citeauthoryear{Kewley, Geller, \& Jansen}{2004}]{Kewley2004} Kewley L.~J., Geller M.~J., Jansen R.~A., 2004, AJ, 127, 2002 

\bibitem[\protect\citeauthoryear{Kitzbichler \& White}{2007}]{KW_millennium2007} Kitzbichler M.~G., White S.~D.~M., 2007, MNRAS, 376, 2 


\bibitem[\protect\citeauthoryear{Kova{\v c} et al.}{2010}]{Kovac2010} Kova{\v c} K., et al., 2010, ApJ, 708, 505 

\bibitem[\protect\citeauthoryear{Kurk et al.}{2009}]{Kurk2009} Kurk J., et al., 2009, A\&A, 504, 331 

\bibitem[\protect\citeauthoryear{Larson, Tinsley, \& Caldwell}{1980}]{Larson1980} Larson R.~B., Tinsley B.~M., Caldwell C.~N., 1980, ApJ, 237, 692 

\bibitem[\protect\citeauthoryear{Leauthaud et al.}{2010}]{Leauthaud2010} Leauthaud A., et al., 2010, ApJ, 709, 97 

\bibitem[\protect\citeauthoryear{Lewis et al.}{2002}]{Lewis2002} Lewis I., et al., 2002, MNRAS, 334, 673

\bibitem[\protect\citeauthoryear{Lilly et al.}{2007}]{Lillyetal2007} Lilly S.~J., et al., 2007, ApJS, 172, 70 

\bibitem[\protect\citeauthoryear{Lutz et al.}{2010}]{Lutz2010} Lutz D., et al., 2010, ApJ, 712, 1287 

\bibitem[\protect\citeauthoryear{Magnelli et al.}{2011}]{magnelli11} Magnelli B., Elbaz D., Chary R.~R., Dickinson M., Le Borgne D., Frayer D.~T., Willmer C.~N.~A., 2011, A\&A, 528, A35 

\bibitem[\protect\citeauthoryear{Magnelli et al.}{2013}]{magnelli2013} Magnelli B., et al., 2013, arXiv, arXiv:1303.4436 

\bibitem[\protect\citeauthoryear{McGee et al.}{2009}]{McGee2009} McGee S.~L., Balogh M.~L., Bower R.~G., Font A.~S., McCarthy I.~G., 2009, MNRAS, 400, 937 


\bibitem[\protect\citeauthoryear{Moore et al.}{1996}]{Moore1996} Moore B., Katz N., Lake G., Dressler A., Oemler A., 1996, Natur, 379, 613 

\bibitem[\protect\citeauthoryear{Muldrew et al.}{2012}]{Muldrew2012} Muldrew S.~I., et al., 2012, MNRAS, 419, 2670 

\bibitem[\protect\citeauthoryear{Muzzin et al.}{2012}]{Muzzin2012} Muzzin A., et al., 2012, ApJ, 746, 188 

\bibitem[\protect\citeauthoryear{Netzer et al.}{2007}]{Netzer2007} Netzer H., et al., 2007, ApJ, 666, 806 

\bibitem[\protect\citeauthoryear{Noeske et al.}{2007a}]{Noeske2007a} Noeske K.~G., et al., 2007a, ApJ, 660, L43 
\bibitem[\protect\citeauthoryear{Noeske et al.}{2007b}]{Noeske2007b} Noeske K.~G., et al., 2007b, ApJ, 660, L47 

\bibitem[\protect\citeauthoryear{Peng et al.}{2010}]{Peng2010} Peng Y., et al., 2010, ApJ, 721, 193 

\bibitem[\protect\citeauthoryear{Peng et al.}{2012}]{Peng2011} Peng Y.-j., Lilly S.~J., Renzini A., Carollo M., 2012, ApJ, 757, 4 
\bibitem[\protect\citeauthoryear{Poglitsch et al.}{2010}]{poglitsch10} Poglitsch, A., et al. 2010, A\&A, 518, L2 

\bibitem[\protect\citeauthoryear{Popesso et al.}{2009}]{Popesso2009} Popesso P., et al., 2009, A\&A, 494, 443 

\bibitem[\protect\citeauthoryear{Popesso et al.}{2011}]{popesso11} Popesso P., et al., 2011, A\&A, 532, A145 
\bibitem[\protect\citeauthoryear{Popesso et al.}{2012}]{popesso2012} Popesso P., et al., 2012, A\&A, 537, A58 

\bibitem[\protect\citeauthoryear{Porter \& Raychaudhury}{2007}]{Porter_Raychaydhury2007} Porter S.~C., Raychaudhury S., 2007, MNRAS, 375, 1409 
\bibitem[\protect\citeauthoryear{Presotto et al.}{2012}]{Presotto2012} Presotto V., et al., 2012, A\&A, 539, A55 


\bibitem[\protect\citeauthoryear{Rasmussen et al.}{2012}]{Rasmussen2012} Rasmussen J., Mulchaey J.~S., Bai L., Ponman T.~J., Raychaudhury S., Dariush A., 2012, ApJ, 757, 122 

\bibitem[\protect\citeauthoryear{Rettura et al.}{2010}]{Rettura2010} Rettura A., et al., 2010, ApJ, 709, 512 

\bibitem[\protect\citeauthoryear{Rodighiero et al.}{2010}]{rodighiero10} Rodighiero, G., et al., 2010, A\&A, 518, L25

\bibitem[\protect\citeauthoryear{Rosario et al.}{2013}]{Rosario2013} Rosario D.~J., et al., 2013, arXiv, arXiv:1302.1202 

\bibitem[\protect\citeauthoryear{Santini et al.}{2012}]{Santini2012} Santini P., et al., 2012, A\&A, 540, A109 

\bibitem[\protect\citeauthoryear{Scodeggio et al.}{2008}]{Scodeggio2008} Scodeggio M., Franzetti P., Garilli B., Fumana M., Paioro L., Zanichelli A., 2008, eic..work, 95 

\bibitem[\protect\citeauthoryear{Scodeggio et al.}{2009}]{Scodeggio2009} Scodeggio M., et al., 2009, A\&A, 501, 21 

\bibitem[\protect\citeauthoryear{Scoville et al.}{2007}]{Scoville2007} Scoville N., et al., 2007, ApJS, 172, 1 

\bibitem[\protect\citeauthoryear{Shao et al.}{2010}]{Shao2010} Shao L., et al., 2010, A\&A, 518, L26 

\bibitem[\protect\citeauthoryear{Silverman et al.}{2010}]{Silverman2010} Silverman J.~D., et al., 2010, ApJS, 191, 124 

\bibitem[\protect\citeauthoryear{Springel et al.}{2005}]{Springel2005} Springel V., et al., 2005, Natur, 435, 629 

\bibitem[\protect\citeauthoryear{Strateva et al.}{2001}]{Strateva2001} Strateva I., et al., 2001, AJ, 122, 1861 

\bibitem[\protect\citeauthoryear{Tanaka et al.}{2009}]{Tanaka2009} Tanaka M., Lidman C., Bower R.~G., Demarco R., Finoguenov A., Kodama T., Nakata F., Rosati P., 2009, A\&A, 507, 671 

\bibitem[\protect\citeauthoryear{Tanaka et al.}{2012}]{Tanaka2012} Tanaka M., et al., 2012, PASJ, 64, 22 


\bibitem[\protect\citeauthoryear{Tran et al.}{2010}]{Tran2010} Tran K.-V.~H., et al., 2010, ApJ, 719, L126 

\bibitem[\protect\citeauthoryear{Whitaker et al.}{2012}]{Whitaker2012} Whitaker K.~E., van Dokkum P.~G., Brammer G., Franx M., 2012, ApJ, 754, L29 



\bibitem[\protect\citeauthoryear{Wilman et al.}{2008}]{Wilman2008} Wilman D.~J., et al., 2008, ApJ, 680, 1009 

\bibitem[\protect\citeauthoryear{Wilman, Zibetti, \& Budav{\'a}ri}{2010}]{Wilman2010} Wilman D.~J., Zibetti S., Budav{\'a}ri T., 2010, MNRAS, 406, 1701 




\bibitem[\protect\citeauthoryear{Woo et al.}{2013}]{Woo2012} Woo J., et al., 2013, MNRAS, 428, 3306 

\bibitem[\protect\citeauthoryear{Wuyts et al.}{2011}]{Wuyts2011} Wuyts S., et al., 2011, ApJ, 742, 96 

\bibitem[\protect\citeauthoryear{Xue et al.}{2011}]{Xue2011} Xue Y.~Q., et al., 2011, ApJS, 195, 10 

\bibitem[\protect\citeauthoryear{Yang et al.}{2007}]{Yang2007} Yang X., Mo H.~J., van den Bosch F.~C., Pasquali A., Li C., Barden M., 2007, ApJ, 671, 153 

\bibitem[\protect\citeauthoryear{York et al.}{2000}]{York2000} York D.~G., et al., 2000, AJ, 120, 1579 

\bibitem[\protect\citeauthoryear{Zabludoff \& Mulchaey}{1998}]{Zabludoff_and_Mulchaey_1998} Zabludoff A.~I., Mulchaey J.~S., 1998, ApJ, 496, 39 

\bibitem[\protect\citeauthoryear{Ziparo et al.}{2013}]{Ziparo2013} Ziparo F., et al., 2013, MNRAS, 434, 3089 


\end{thebibliography}
\end{document}